\documentclass[acmsmall,screen]{acmart}

\pdfoutput=1

\usepackage{listings}
\usepackage{subcaption}
\usepackage{graphicx}
\usepackage{wrapfig}
\usepackage{makecell}
\usepackage{svg}
\usepackage{enumitem}
\usepackage{amsmath}
\usepackage{multirow}

\usepackage{enumitem}
\usepackage{xspace}
\usepackage{pifont}
\usepackage{diagbox}
\usepackage{multirow}
\usepackage{float}
\usepackage{makecell}
\usepackage{tabularx}
\usepackage{listings}
\usepackage{tikz}
\usepackage{pgf-pie}
\usepackage{subcaption}
\usepackage{setspace}
\usepackage[vlined,ruled,linesnumbered]{algorithm2e}
\usepackage{tcolorbox}
\usepackage{tablefootnote}
\usepackage{wrapfig}
\usepackage{pifont}

\SetKw{Input}{Input}
\SetKw{Output}{Output}

\AtBeginDocument{%
  }

\setcopyright{acmlicensed}
\copyrightyear{2025}
\acmYear{2025}
\acmDOI{XXXXXXX.XXXXXXX}
\acmConference[Conference acronym 'XX]{Make sure to enter the correct
  conference title from your rights confirmation email}{June 03--05,
  2018}{Woodstock, NY}
\acmISBN{978-1-4503-XXXX-X/2018/06}

\begin{document}

\title{Type-aware LLM-based Regression Test Generation for Python Programs}

\author{Runlin Liu}
\affiliation{
  \institution{Beihang University}
  \country{China}
}
\email{runlin22@buaa.edu.cn}

\author{Zhe Zhang}
\affiliation{
  \institution{Beihang University}
  \country{China}
}
\email{zhangzhe2023@buaa.edu.cn}

\author{Yunge Hu}
\affiliation{
  \institution{Beihang University}
  \country{China}
}
\email{hygchn04@gmail.com}

\author{Yuhang Lin}
\affiliation{
  \institution{Beihang University}
  \country{China}
}
\email{yuhanglin35@gmail.com}

\author{Xiang Gao}
\affiliation{
  \institution{Beihang University}
  \country{China}
}
\email{xiang_gao@buaa.edu.cn}

\author{Hailong Sun}
\affiliation{
    \institution{Beihang University}
    \country{China}
}
\email{sunhl@buaa.edu.cn}

  \newcommand{\nbc}[3]{
        {\colorbox{#3}{\scriptsize\textcolor{white}{#1}}}
            {\textcolor{#3}{\small$\blacktriangleright$\textit{#2}$\blacktriangleleft$}}
}

\newcommand{\change}[1]{{\color{red}#1}}

\newcommand{\xiang}[1]{\nbc{XG}{#1}{blue}}
\newcommand{\zhangzhe}[1]{\nbc{ZZ}{#1}{cyan}}
\newcommand{\rl}[1]{\nbc{RL}{#1}{green}}
\newcommand{\yuhang}[1]{\nbc{YH}{#1}{orange}}
\newcommand{\yunge}[1]{\nbc{YG}{#1}{purple}}

\newcommand{\toolname}[0]{\textsc{Test4Py}\xspace}

\newcommand{\todo}[1]{{\color{yellow}#1}}

\begin{abstract}
Automated regression test generation has been extensively explored, yet generating high-quality tests for Python programs remains particularly challenging.
Because of the Python's dynamic typing features, existing approaches, ranging from search-based software testing (SBST) to recent LLM-driven techniques, are often prone to type errors.
Hence, existing methods often generate invalid inputs and semantically inconsistent test cases, which ultimately undermine their practical effectiveness.
To address these limitations, we present \toolname, a novel framework that enhances type correctness in automated test generation for Python.
\toolname leverages the program’s call graph to capture richer contextual information about parameters, and introduces a behavior-based type inference mechanism that accurately infers parameter types and construct valid test inputs. 
Beyond input construction, \toolname integrates an iterative repair procedure that progressively refines generated test cases to improve coverage.
In an evaluation on 183 real-world Python modules, \toolname achieved an average statement coverage of 83.0\% and branch coverage of 70.8\%, outperforming state-of-the-art tools by 7.2\% and 8.4\%, respectively. 
\end{abstract}




\maketitle

\section{Introduction}
\label{sec:intro}

Regression testing is a cornerstone of modern software engineering, playing a critical role in ensuring software correctness and sustaining long-term code quality. 
However, the manual construction of regression tests remains labor-intensive and error-prone~\cite{Survey_of_coverage}, which has spurred extensive research on automated test generation. 
Classical approaches encompass random-based techniques~\cite{davis2023nanofuzz, wei2022free}, constraint-driven strategies~\cite{hwang2021justgen, godefroid2005dart, ma2015grt}, and search-based software testing (SBST)~\cite{zhou2022selectively, andrews2011genetic, Codamosa, lukasczyk2022pynguin}. 
Recent advances in Large Language Models (LLMs) have further energized this field, as illustrated by CodaMosa~\cite{Codamosa}, which augments SBST with LLM-generated tests, and CoverUp~\cite{CoverUp}, which employs coverage-guided prompting to progressively steer LLMs toward higher-coverage test suites.

Despite these encouraging developments, automated test generation remains particularly challenging for dynamically typed languages such as Python.
In contrast to statically typed languages, Python omits mandatory type annotations, which obscures parameter types and class affiliations. 
Empirical studies underscore the severity of this limitation: nearly 30\% of developer-reported issues on GitHub and Stack Overflow arise from type-related errors~\cite{oh2022pyter}. 
Since the validity and semantic soundness of inputs are highly important in test cases, type correctness is indispensable for generating meaningful and effective test cases.
As discussed in prior work~\cite{lukasczyk2022pynguin}, consider the function $triangle(a,b,c)$ intended to determine whether three sides form an equilateral triangle. 
While the parameters are expected to be \textbf{int} or \textbf{float}, Python’s permissive typing allows structurally valid but semantically irrelevant inputs, such as lists of strings.
Such cases may execute without failure and even attain full code coverage, yet they are semantically meaningless, provide no practical regression protection, and ultimately undermine the utility of the generated tests.

Unfortunately, existing approaches exhibit limited accuracy in generating tests with correct types. 
This deficiency arises from two key factors: (i) parameter construction often lie outside function bodies, depriving models of essential local context; and (ii) user-defined types are highly project-specific and sparsely represented in pretraining corpora, severely restricting model generalization. 
These limitations crystallize into a fundamental research gap: the lack of mechanisms to generate semantically valid test inputs for dynamically typed languages.

This gap motivates the following research questions:  
\begin{itemize}[leftmargin=*]
    \item How can parameter types be accurately inferred for Python functions in the absence of explicit type annotations and under the constraints of dynamic typing?  
    \item How can program context be effectively constructed and exploited to guide LLMs in synthesizing valid and semantically meaningful test inputs, particularly for complex user-defined types?  
\end{itemize}

\textbf{\textit{Our approach.}}
To address these challenges, we present \toolname, a framework that infers function parameter types prior to test case generation, thereby enhancing the type correctness of the generated test inputs.
Although Python’s dynamic typing provides considerable flexibility, it also introduces a fundamental obstacle to precise type inference.
To overcome this limitation, \toolname draws inspiration from the classic ``Duck Test''\footnote{\url{https://en.wikipedia.org/wiki/Duck_test}}, and introduces behavior-guided parameter inference (BGPI). 
BGPI conceptualizes type resolution the process of identifying a parameter’s type through its observable behaviors at both the syntactic and semantic levels.
To do so, LLMs must accurately capture the semantic roles of parameters. 
However, parameters in real-world software rarely exist in isolation; they interact through intricate call relationships spanning multiple modules.
To model such dependencies, we construct a project-level call graph and derive parameter-centric semantic summaries from both callee behaviors and caller contexts, thereby enabling context-aware semantic inference.

Subsequently, \toolname leverages the inferred types to construct a compact but expressive \textit{type context}.
This enriched representation guides LLMs in synthesizing executable and semantically valid test inputs. 
To further improve robustness, \toolname introduces an adaptive error repair mechanism that autonomously performs error resolution.

We evaluate \toolname on 183 real-world Python modules, and compare it with CodaMosa~\cite{Codamosa} and CoverUp~\cite{CoverUp}. 
Evaluation result show that \toolname achieves an average statement coverage of 83.0\% and branch coverage of 70.8\%, surpassing state-of-the-art baselines by 7.2\% and 8.4\%, respectively.
An ablation study further reveals that the proposed type inference component alone improves coverage by 13.5\% in settings lacking explicit type annotations.

In summary, this paper makes the following contributions:
\begin{itemize}[leftmargin=*]
    \item We propose a novel approach for parameter-centric summarization that leverages a combination of callee-driven behaviors and caller-contextual evidence to enhance the type-correctness and semantic-correctness of generated test cases.
    \item We propose a behavior-guided type inference framework that integrates syntactic filtering with semantic retrieval to enhance inference accuracy, and further exploits the inferred type information to enhance the quality of LLM-based Python test case generation. We have released the source code of \toolname on \url{https://github.com/Test4DT/Test4Py}.
    \item We conduct a comprehensive evaluation on real-world projects, demonstrating significant improvements in statement and branch coverage over state-of-the-art baselines.  
\end{itemize}

\section{Motivation}

We use the \texttt{get\_custom\_loader} function from the \texttt{PyCG} \footnote{PyCG: \url{https://github.com/vitsalis/PyCG}} project 
as a representative example to illustrate our motivation. 
\texttt{PyCG} is an open-source project on GitHub with 332 stars.
It generates call graphs by analyzing Python code and supports advanced features such as higher-order functions and complex class inheritance structures.

\begin{figure}[htp]
\begin{minipage}[t]{0.50\textwidth}
\centering
\begin{lstlisting}[numbers=none]{python}
def get_custom_loader(ig_obj):
  class CustomLoader(importlib.abc.SourceLoader):
    def __init__(self, fullname, path):
      self.fullname = fullname
      self.path = path
      ig_obj.create_edge(self.fullname)
      if not ig_obj.get_node(self.fullname):
        ig_obj.create_node(self.fullname)
        ig_obj.set_filepath(self.fullname, self.path)
    <...omitted code...>
  return CustomLoader
\end{lstlisting}
\vspace{-8pt}
\subcaption{Part of the get\_custom\_loader function}
\label{motivation-source-code}
\begin{lstlisting}[numbers=none]{python}
def test_custom_loader():
  ig_obj=MockImportGraph()
  loader=get_custom_loader(ig_obj)
  <...omitted code...>
\end{lstlisting}
\vspace{-8pt}
\subcaption{Test case Generated by LLM}
\label{motivation-CodaMosa}
\end{minipage}
\hfill
\begin{minipage}[t]{0.46\textwidth}
\centering
\begin{lstlisting}[numbers=none]{python}
def test_case_4():
  try:
    bool_0 = True
    var_0 = module_0.get_custom_loader(bool_0)
    <...omitted code...>
  except BaseException:
    pass
\end{lstlisting}
\vspace{-8pt}
\subcaption{Test case Generated by SBST}
\label{motivation-SBST}
\begin{lstlisting}[numbers=none]{python}
class ImportManager(object): 
  def create_node(self, name):
    self.import_graph[name] = {
    "filename":"", "imports":set()}

  def set_filepath(self, node_name, filename):
    node = self.get_node(node_name)
    node["filename"] = os.path.abspath(filename)
\end{lstlisting}
\vspace{-8pt}
\subcaption{Part of the ImportManager class}
\label{motivation-ImportManager}
\end{minipage}
\captionsetup{type=listing}
\vspace{-8pt}
\captionof{lstlisting}{An example of code and the corresponding auto-generated unit tests.
}
\label{Codes-Motivation-1}
\vspace{-8pt}
\end{figure}

As shown in Listing~\ref{motivation-source-code}, the function \texttt{get\_custom\_loader} returns a module loader \texttt{CustomLoader}. 
During initialization, the parameter \texttt{ig\_obj}, which is an instance of the \texttt{ImportManager} class presented in Listing~\ref{motivation-ImportManager}, is responsible for maintaining the relationships among imported modules. 
Specifically, it first establishes the import edges and subsequently verifies whether the target module already exists in the import graph. 
If the module is not present, a new node is created, and the corresponding file path is recorded.

We employ the unit test generation tool CodaMosa \cite{Codamosa} to generate test cases for the function \texttt{get\_custom\_loader}. 
Initially, CodaMosa utilizes search-based software testing (SBST) approach to produce simple test cases, as shown in Listing~\ref{motivation-SBST}. 
This test case sets \texttt{bool\_0} to \texttt{True} and passes it to \texttt{get\_custom\_loader}.
However, since the constructor of \texttt{CustomLoader} accesses members of \texttt{ig\_obj}, when \texttt{ig\_obj} is initialized as boolean, the execution of \texttt{CustomLoader}'s instantiation will result in a runtime error, preventing further improvements in coverage.
When SBST fails to enhance coverage, CodaMosa activates the large language model to generate test cases.
One of the test cases generated by LLM is shown in Listing~\ref{motivation-CodaMosa}.
In this case, LLM mocks a class named \texttt{MockImportGraph} and uses its instance as the parameter of \texttt{get\_custom\_loader}.
\texttt{MockImportGraph} class ensures that all members accessed by the \texttt{\_\_init\_\_} function of \texttt{CustomLoader} are present, thus preventing run-time errors.
However, due to the absence of explicit type information for \texttt{ig\_obj}, LLM is unable to infer the exact implementations of methods such as \texttt{create\_node}.
Instead, it relies solely on method names to approximate their behavior, leading to inconsistencies between the generated test case and the actual implementation.
CodaMosa then makes 16 additional attempts, and all of which fail.
The primary cause of this failure is that CodaMosa does not provide type information in its prompts, preventing LLM from generating semantically valid test cases.

\begin{wrapfigure}{r}{0.60\textwidth}
\begin{lstlisting}[numbers=none]{python}
class TestGetCustomLoader(unittest.TestCase):
  def setUp(self):
    self.ig_obj = ImportManager()
    self.loader_class = get_custom_loader(self.ig_obj)

  def test_loader(self):
    loader = self.loader_class("test.module", 
      "/path/to/module.py")
    self.assertIn("test.module", self.ig_obj.import_graph)
    <...omitted code...>
\end{lstlisting}
\captionsetup{type=listing}
\vspace{-6pt}
\captionof{lstlisting}{Test case generate by \toolname}
\label{motivation-us}
\vspace{-10pt}
\end{wrapfigure}

To infer the type of \texttt{ig\_obj}, we employ Hityper~\cite{peng2022static}, which is the state-of-the-art tool for type inference.
However, Hityper incorrectly infers \texttt{ig\_obj} as a \texttt{str} type.
This incorrect inference arises because \texttt{ig\_obj} is a function parameter without any direct assignment statements to aid type inference.
Moreover, since \texttt{ig\_obj} is a user-defined type, the deep learning component of Hityper struggles to infer its type accurately.

In contrast, our proposed \toolname employs a type inference mechanism to determine the type of the argument \texttt{ig\_obj}.
In the \texttt{\_\_init\_\_} function of \texttt{CustomLoader}, \texttt{ig\_obj} invokes member methods such as \texttt{create\_edge} and \texttt{get\_node}, which are responsible for maintaining module import relationships.
To resolve the type of \texttt{ig\_obj}, \toolname identifies classes that define these methods and are semantically related to import management.
It subsequently locates the \texttt{ImportManager} class and extracts information about it.
By incorporating this information into the prompt, LLM can instantiate an \texttt{ImportManager} object and pass it as an argument to \texttt{get\_custom\_loader}, as demonstrated in Listing~\ref{motivation-us}.
This approach enables \toolname to successfully instantiate \texttt{loader\_class} and achieve full statement coverage for \texttt{get\_custom\_loader}, thereby not only improving test coverage but also generating assertions that verify the core functionality of the function.

This example demonstrates that accurately inferring type information for key variables can significantly enhance both the coverage of generated test cases and the quality of assertions.
Existing approaches, when handling dynamically typed languages like Python, frequently overlook critical contextual information, resulting in poor-quality test cases generated by LLM.
Addressing this limitation requires a method capable of searching across the entire project, filtering relevant information, and accurately inferring variable types.
\toolname fulfills this requirement, making it a promising solution for improving automated test generation in dynamically typed languages.







       

\section{Methodology}

This section presents \toolname, a framework designed to advance automated unit test generation for Python programs by systematically addressing the challenges posed by dynamic typing and limited type annotations.

\subsection{Framework Design}
\label{subsec:framework}

\begin{figure}[t!]
\centering
\includegraphics[width=1\textwidth]{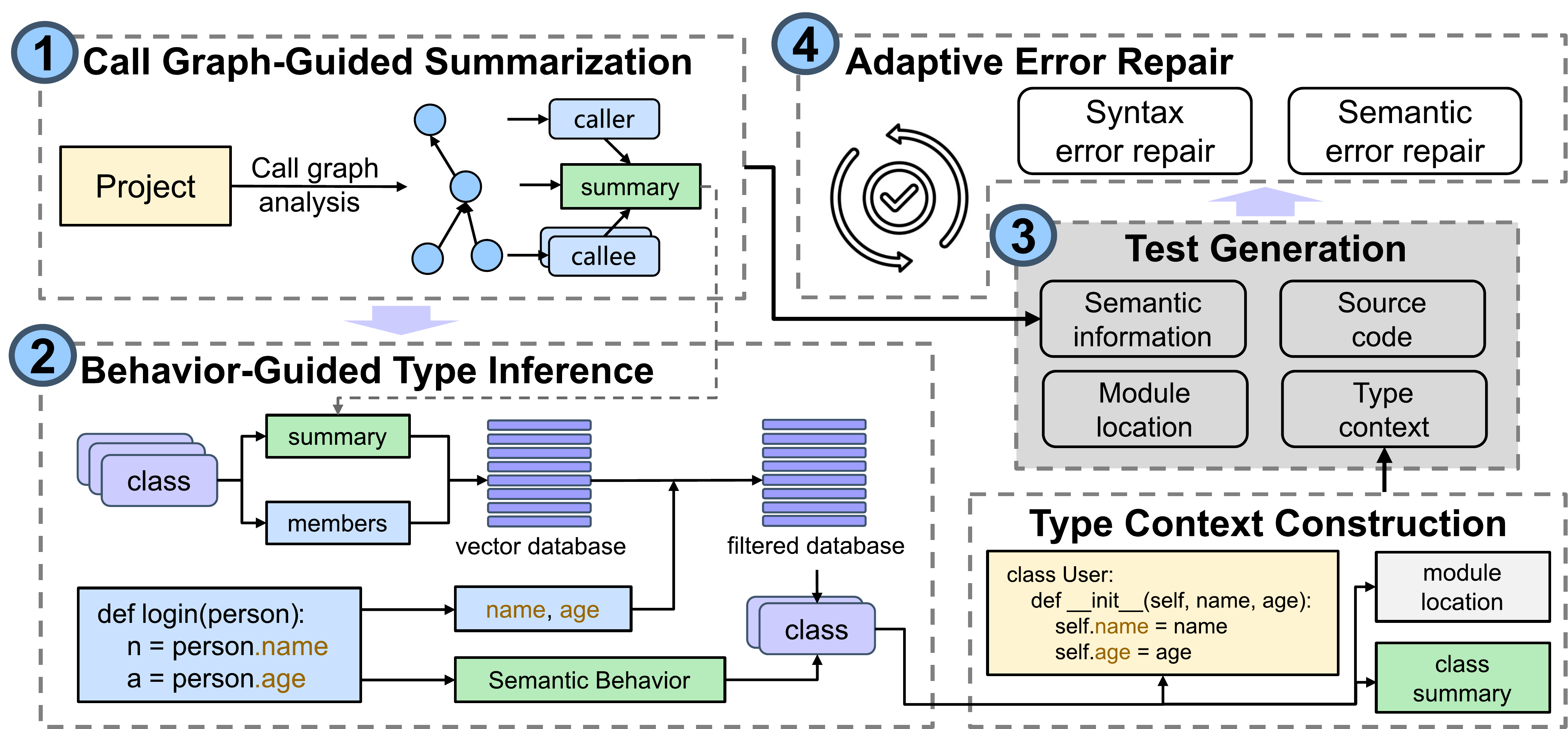}
\caption{The overall architecture of \toolname, comprising four interdependent stages: call graph-guided parameter-centric summarization, behavior-guided parameter type inference, type-guided test case generation, adaptive error repair.}
\label{fig:overall}
\end{figure}

As illustrated in Figure~\ref{fig:overall}, the workflow of \toolname is structured into four tightly coupled stages, each addressing a fundamental obstacle in unit test generation for dynamically typed languages,
where explicit type annotations are scarce and semantic relations among program entities are often implicit.

First of all, to enhance the semantic comprehension of function parameters, \toolname first constructs a project-wide call graph at stage \ding{172}. 
This structural representation enables the derivation of parameter-centric semantic summaries, capturing how each function interacts with its surrounding context. 
These summaries provide enriched contextual knowledge that significantly improves the accuracy of subsequent inference steps. 
Subsequently, building upon the function-level summaries,  at stage \ding{173}, \toolname derives class-level semantic summaries by aggregating the behaviors of their constituent methods. 
These class summaries are then embedded and stored in a vector database, forming the foundation for whole-project type inference.
For each function parameter, syntactic behavior constraints are applied to eliminate incompatible candidates, after which semantic behavior retrieval aligns the observed parameter usage with the most plausible classes.
This explicit coupling of structural and semantic evidence goes beyond conventional inference approaches, yielding type predictions that are both precise and contextually grounded.
Then, once candidate classes are inferred, at stage \ding{174}, \toolname constructs a comprehensive type context that integrates the class constructor, module path, and functional summary. 
This enriched context serves as the foundation for LLM-based test generation, guiding the model to synthesize more faithful and executable test cases. 
To further enhance robustness, stage \ding{175} incorporates an adaptive error-repair mechanism that automatically detects and corrects both syntactic flaws and semantic inconsistencies in the generated tests. 

\subsection{Call Graph-Guided Parameter-Centric Summarization}
\label{subsec:summary}

\noindent
\textbf{Motivation.}  
Recent studies have shown that LLM-generated natural language summaries of functions can enhance automated test case generation~\cite{yuan2023no}, as they provide richer semantic representations of the functions under test.  
However, existing function summarization approaches are not designed for test generation task.
For test generation, the most critical semantic unit is the function parameter. 
Parameters embody both the interface contract and the input-output behaviors of software components, and inaccuracies in parameter understanding directly compromise the validity of generated tests.  
We therefore advocate a parameter-centric perspective: our goal is to produce summaries that explicitly model parameter types, constraints, and usage intents.
Yet, parameters in real-world functions are rarely self-contained; their semantics emerge through complex inter-procedural interactions, where callees constrain permissible parameter values and callers project realistic usage patterns.
Without capturing these dependencies, LLM-generated summaries risk being incomplete or misleading.


\noindent
\textbf{Approach.}  
To address this challenge, \toolname introduces a \emph{Call Graph-Guided Parameter-Centric Summarization} framework, where project-level call graph analysis serves as the structural backbone to integrate two complementary sources of parameter semantics:  
(i) \emph{callee-driven behavioral constraints}, which reveal how parameter values influence downstream computations; and  
(ii) \emph{caller-contextual evidence}, which illuminates the concrete types and intents associated with parameter usage.  
We adopt \texttt{pycg}~\cite{salis2021pycg}, a state-of-the-art static call graph construction framework, to ensure reliable extraction of inter-procedural dependencies.

\subsubsection{Callee-Driven Behavioral Analysis}

This phase focuses on how parameter propagate effects across the call chain.  
The summary of a callee not only provides the behavioral semantics represented by the call expressions but also reveals the relational semantics among its parameters.
For example, if a callee checks whether two arguments are equal, this constraint informs that the parameters may satisfy an equality condition.
Capturing such parameter-centric constraints requires analyzing not only immediate callees but also deeper invocations along the call chain.
A simple traversal of direct function calls is therefore insufficient, as it would omit semantic information carried through deeper dependencies.
To overcome this limitation, we adopt a topological sorting approach, where each function’s summarization incorporates the summaries of the functions it calls.
This method allows us to capture and refine parameter-related semantics across the entire call chain.

Formally, let $Called(f)$ denote the set of functions invoked by $f$, and $f.sc$ its source code. Given an instruction $be\_instruction$, which directs LLM to analyze the roles of function parameters and how modifications to their values influence the execution of the function
, the forward topological summary of $f$ is defined as:

\begin{equation}
f.\mathit{be\_summary} =
\operatorname{LLM}\big(be\_instruction + f.sc + \sum_{x \in Called(f)} x.\mathit{be\_summary}\big).
\end{equation}

To handle recursion, cycles in the call graph are temporarily relaxed by removing one dependency edge, ensuring termination while preserving semantic integrity.

\subsubsection{Caller-Contextual Semantic Projection}

Although forward traversal effectively captures the fine-grained semantics of callees, it inherently overlooks the \emph{usage-driven} semantics that arise from domain-specific conventions, how developers in a particular application domain consistently interpret and utilize functions within conventional usage contexts.
For example, a function invocation \textit{triangle(3, 4, 5)} reflect the semantics of determining whether three integers compose a triangle,
implying that its parameters represent numeric types such as \texttt{int} or \texttt{float}.
To address this, we introduce a \emph{caller-contextual summarization} strategy that traverses the call graph in reverse topological order, treating each function as a semantic construct shaped by the behavioral constraints imposed by its callers and their domain-specific usage conventions.

This strategy fulfills two complementary objectives:
\begin{itemize}[leftmargin=*]
    \item \textbf{Parameter Type Inference.} Caller-contextual evidence enables more accurate preliminary parameter type inference by leveraging concrete invocation patterns.
    For example, if a caller invokes a function as \texttt{add(1, 2)}, LLM can infer that both parameters are likely of type \texttt{int}.
    \item \textbf{Semantic Intent Modeling.} Beyond type inference, caller-contextual projection facilitates the inference of semantic intent. 
    This helps LLM in generating test cases that are better aligned with realistic usage scenarios and domain-specific conventions.
\end{itemize}

Moreover, for functions with zero in-degree in the call graph (i.e., top-level entry points), \toolname augments the summarization process with auxiliary resources such as project-level documentation (e.g., \texttt{README.md}).
This supplementation establishes domain-level semantics that cannot be derived solely from intra-project call relationships.

Formally, let $Call(f)$ denote the set of functions invoking $f$, and $docs$ denote the project-level documentation. 
Given an instruction $se\_instruction$, which directs LLM to analyze the semantic roles and types of function parameters, the reverse summary is defined as:
\begin{equation}
f.\mathit{se\_summary} =
\begin{cases}
\operatorname{LLM}(se\_instruction + x.sc + x.\mathit{se\_summary}), & Call(f) \neq \emptyset,~ x \in Call(f)\\
\operatorname{LLM}(se\_instruction + docs), & Call(f) = \emptyset.
\end{cases}
\end{equation}

Due to the limited context window of LLMs, it is impractical to include all elements of $Call(f)$ in the prompt.
Consequently, certain semantic cues from unobserved callers may be omitted, leading to an incomplete $se\_summary$.
To address this limitation, the generated $se\_summary$ is treated as an auxiliary semantic reference, rather than as a strict constraint on semantics in subsequent stages.

\subsubsection{Summary Synthesis}

Finally, \toolname synthesizes $f.\mathit{be\_summary}$ and $f.\mathit{se\_summary}$ to produce the final summary $f.\mathit{summary}$. 
This synthesis step removes redundant information and aligns fine-grained behavioral details with higher-level abstractions, resulting in a concise yet comprehensive semantic representation that mitigates redundancy and prevents information overload in subsequent steps.

\noindent
\textbf{Example.} 
To demonstrate how our approach integrates behavioral and semantic cues for precise parameter understanding, we present a representative example. 
\autoref{list:sc_transform} depicts the function \texttt{transform} along with its caller and callee. 
The function aims to scale all \texttt{Record} objects contained in \texttt{x} by a factor of $y$.

\begin{figure}[htp]
\begin{minipage}[t]{0.48\textwidth}
\centering
\begin{lstlisting}[numbers=none]{python}
def transform(x, y):
    results = process_all(x, y)
    <...omitted code...>

def process_all(objs, scale_factor):
    for o in objs:
        o.scale(scale_factor)
    return [o.average() for o in objs]

def test():
    r1 = Record([1, 2, 3])
    r2 = Record([10, 20])
    transform([r1, r2], 2.5)
\end{lstlisting}
\vspace{-6pt}
\subcaption{The function \texttt{transform} and its related functions.}
\label{list:sc_transform}
\end{minipage}
\hfill
\begin{minipage}[t]{0.48\textwidth}
\centering
\begin{lstlisting}[numbers=none]{}
x (list): A list of objects to be transformed. Each object must have `scale` and `average` methods.
y (float): The scaling factor used to transform the objects.
\end{lstlisting}
\vspace{-6pt}
\subcaption{Behavioral summary ($be\_summary$) of \texttt{transform}.}
\vspace{6pt}
\label{list:be_summary}
\begin{lstlisting}[numbers=none]{}
x (List[Record]): A list of Record objects.
y (float): A scaling factor used to scale each Record object in the list.
\end{lstlisting}
\vspace{-6pt}
\subcaption{Semantic summary ($se\_summary$) of \texttt{transform}.}
\label{list:se_summary}
\end{minipage}
\vspace{-6pt}
\caption{Example demonstrating how \toolname synthesizes behavioral and semantic summaries.}
\vspace{-6pt}
\end{figure}

The \emph{callee-driven behavioral analysis} exploits the summary of \texttt{process\_all}, which encodes its iteration and scaling behaviors over \texttt{Record} objects. 
From this behavioral evidence, the model infers that the parameter $x$ must support both \texttt{scale} and \texttt{average} methods, as captured in the $be\_summary$ (\autoref{list:be_summary}). 
Subsequently, the \emph{caller-contextual semantic projection} analyzes the invocation context of \texttt{transform} in \texttt{test}, where the concrete arguments reveal that $x$ is of type \texttt{List[Record]}, as shown in the $se\_summary$ (\autoref{list:se_summary}). 

However, due to the dynamic language features inherent to Python, no single static analysis algorithm can generate a perfect call flow graph. Even the robust pycg framework exhibits a recall rate of only 69.9\%. 
This fundamental limitation indicates that relying solely on the Call Graph-Guided summarization method is insufficient to fully resolve the parameter type inference problem in Python programs.

\subsection{Behavior-Guided Parameter Type Inference} 
\label{subsec:type-infer}

As highlighted in Section~\ref{sec:intro}, accurately resolving the types of dynamically defined function parameters remains a fundamental obstacle to automated test case generation. 
The challenge is particularly acute for user-defined types with intricate internal structures, where traditional inference strategies fail to capture the semantic dependencies between program entities and their contextual usage.  
Existing methods either depend on incomplete annotations or rely on shallow syntactic heuristics, both of which exhibit limited effectiveness in dynamically typed, real-world software systems.

To address this limitation, we introduce the principle of \emph{behavior-guided inference}. 
The central hypothesis is that the observable operational behaviors of a parameter—such as the members it accesses and the operations it supports—provide the most discriminative evidence for type identification.  
This perspective resonates with the intuition underlying the classical ``Duck Test''\footnote{\url{https://en.wikipedia.org/wiki/Duck_test}}, which informally asserts that an entity can be identified by its characteristic behaviors. 
We reformulate type inference as an alignment problem between behavioral evidence and candidate type definitions, thereby extending the heuristic notion into a principled retrieval-augmented framework.
Building on this insight, we propose the \emph{Behavior-Guided Parameter Inference (BGPI)} framework, which integrates syntactic behavior filtering with semantic behavior retrieval to achieve precise and context-aware type resolution.


The overall workflow is summarized in Algorithm~\ref{alg:type_resolution}. 
Parameters are categorized into three groups:  
(1) built-in Python types (e.g., \texttt{int}, \texttt{List}),  
(2) third-party library types (e.g., \texttt{ast.FunctionDef}), and  
(3) project-specific user-defined types.  
The first two categories can be reliably inferred by LLMs through prior knowledge acquired from pretraining. 
The central challenge lies in resolving user-defined types, which are inherently underrepresented in training corpora and thus require project-specific reasoning.

\toolname first builds a knowledge base of all project-specific classes.
Specifically, it applies static analysis (Lines~\ref{line:class-analysis}-\ref{line:class-analysis-end}) to extract all accessible member variables and methods for each class. 
To mitigate the limitations of raw code embeddings, each class is further abstracted by an LLM into a concise functional summary (Line~\ref{line:class-summary}), which captures its high-level semantics while filtering out syntactic noise.
These summaries are encoded into semantic vector representations (Line~\ref{line:class-vector}) and stored in a vector database, forming a knowledge base of user-defined types.

At the inference stage, \toolname performs an initial type inference for each parameter lacking explicit annotations (Line~\ref{line:first-infer}).
\toolname performs behavior-guided type inference\emph{BGTI} by leveraging the knowledge base gathered in the preceding steps.
First, it analyzes the function body to identify all members
accessed by each parameter, including field accesses and method invocations (Line~\ref{line:find-members}), and uses these signals to filter candidate classes in the vector database by enforcing syntactic compatibility (Line~\ref{line:filter}).
Subsequently, the refined candidate set is ranked via semantic retrieval, aligning the parameter’s observed usage behaviors with the most plausible user-defined type (Line~\ref{line:retrive}).
These two stages, which we later formalize as \emph{Syntactic Behavior Filtering} and \emph{Semantic Behavior Retrieval}, jointly ensure both precision and generality in type resolution.

\begin{algorithm}[t]
\caption{Behavior-Guided Type Resolution for Parameter}
\label{alg:type_resolution}
\KwIn{Target program $program$}
\KwOut{Type for all parameters}

$vector\_database \gets \emptyset$;\\
$result \gets \emptyset$; \\
\ForEach{$\mathit{class} \in \mathit{program}$}{
    \label{line:class-analysis}
    $\mathit{class.members} \gets$ \texttt{GetMembers}($\mathit{class}$); \\
    \ForEach{$\mathit{super} \in \mathtt{superClass(}\mathit{class})$}{
        \label{line:class-analysis-end}
        $\mathit{class.members} \gets \mathit{class.members} \cup \texttt{GetMembers}(\mathit{super})$; 
    }
    \label{line:class-summary}
    $\mathit{class.summary} \gets$ \texttt{Summarize}($\mathit{class}$); \\
    \label{line:class-vector}
    $\mathit{vector\_database} \gets \mathit{vector\_database}\ \cup$ \texttt{Vectorize}($\mathit{class.summary}$); \\
}
\ForEach{$f \in \mathit{program}$}{
    \ForEach{$\mathit{para} \in \mathit{f.paras}$}{
        \uIf{HasTypeAnnotation($\mathit{para}$)}{
            $\mathit{type} \gets GetTypeAnnotation(\mathit{para})$;
        }
        \Else{
            \label{line:first-infer}
            $\mathit{type} \gets$ \texttt{InferType}($f$, $\mathit{para}$); \: \tcp{Preliminary inference}
            
            \uIf{$\mathit{type}$ is user-defined}{
                \label{line:find-members}
                $\mathit{members} \gets$ \texttt{FindMembers}($\mathit{para}$, $f$); \: \tcp{Members accessed in $f$}
                \label{line:filter}
                $\mathit{filtered} \gets$ \texttt{FilterByMembers}($\mathit{vector\_database}$, $\mathit{members}$);\\
                \label{line:retrive}
                $\mathit{type} \gets$ \texttt{SemanticRetrieve}($f$, $\mathit{para}$, $\mathit{filtered}$);
            }
        }
        $result \gets result \cup \{ (\mathit{para}, \mathit{type}) \}$;
    }
}
\Return $result$
\end{algorithm}

\subsubsection{Syntactic Behavior Filtering}


The \emph{syntactic behavior filtering} exploits fine-grained structural cues observable in the function body to filter out incompatible types. 
Specifically, the parameter usage behaviors, which manifest as (1) field accesses and (2) method invocations, serve as discriminative signals that enable early pruning of infeasible type candidates.
Formally, give a parameter $\mathit{para}$ in function $f$ and a type $\mathit{class}$, let $AccessMembers(\mathit{para})$ denote the set of accessed members and $DefineMembers(\mathit{class})$ denote the members defined or inherited by $\mathit{class}$.
A necessary condition is:
\begin{equation}
\forall x \in AccessMembers(\mathit{para}), \ (x \notin DefineMembers(\mathit{class}) \implies typeof(\mathit{para}) \neq \mathit{class})
\end{equation}
This constraint enables early elimination of incompatible candidate types. 

We implement this mechanism through Abstract Syntax Tree (AST) analysis. 
For each class, we recursively collect all explicitly defined and inherited members to construct $DefineMembers(\mathit{class})$. 
For each parameter, we parse the function body to extract its accessed members to construct $AccessMembers(\mathit{para})$. 
In addition, we incorporate specialized handling for Python’s built-in object members (e.g., \textit{\_\_dict\_\_}, \textit{\_\_class\_\_}) to prevent spurious eliminations. 

\subsubsection{Semantic Behavior Retrieval}

While syntactic filtering effectively reduces the search space, it alone is insufficient for disambiguating user-defined types.  
To achieve finer-grained discrimination, we introduce \emph{semantic behavior retrieval}, which leverages the reasoning capabilities of LLMs to infer the semantic role of a parameter from its functional usage context. 
For example, in a call \texttt{login(student)}, the parameter \texttt{student} is more plausibly associated with a domain-specific entity (e.g., a student object) rather than a primitive type.
To operationalize this idea, the knowledge base of user-defined types is constructed and indexed through \emph{summary-based semantic representations}. 
Unlike raw code embeddings, which often contain syntactic noise and hinder retrieval accuracy~\cite{eibich2024aragog}, these high-level summaries serve as \emph{semantic anchors} that abstract away irrelevant implementation details while retaining discriminative contextual information. 
The summaries are generated with the assistance of pre-computed function-level descriptions ($f.summary$), ensuring that the semantic representation of a class faithfully reflects the behavioral cues embodied in its constituent methods.
During inference, \toolname prompts LLM to synthesize a semantic query that encapsulates the behavioral role of the target parameter $\mathit{para}$ within the enclosing function. 
The query is executed against the filtered vector database using Maximal Marginal Relevance (MMR)~\cite{mao2020multi}, which balances relevance and diversity to produce a ranked list of candidate classes. 
This design not only improves retrieval precision but also mitigates the risk of overfitting to a narrow subset of types.

Moreover, another critical challenge arises from the non-uniqueness of parameter types in Python.
Polymorphism allows multiple subclasses of a parent class to be valid for the same parameter, while Python’s dynamic typing system further admits structurally or semantically heterogeneous alternatives.
Ignoring semantic variability would constrain test generation, thereby limiting execution path exploration and reducing achievable code coverage.
To address this problem, \toolname retains the top-$k$ retrieved classes based on similarity scores, where $k$ is a configurable parameter. 
This relaxation strategically broadens the candidate type space, enhancing the robustness and diversity of generated test cases. 
Overall, semantic behavior retrieval constitutes a pivotal component of our framework, substantially improving the realism of inferred types and strengthening the exploratory capacity of subsequent test generation. 


\subsection{Type-Guided Test Case Generation}
\label{subsec:generation}

While our framework is able to infer parameter types for the function under test, directly using these inferred types as prompts is insufficient to ensure high-quality test generation. 
On the other hand, incorporating the complete source code of all potential parameter-related classes would substantially inflate the prompt length and impair the effectiveness of the attention mechanism, thereby reducing the reliability of the generated outputs. 
This section introduces a concise yet effective approach to leveraging parameter type information, which reduces spurious generations while preserving semantic fidelity.

\subsubsection{Prompt Design for Test Generation}

To achieve a concise yet informative prompt design, we structure the type context around three complementary components, each targeting a critical aspect of parameter utilization in test case generation:
\begin{itemize}[leftmargin=*]
    \item \textbf{Module Location.} Specifies the module path of the class, ensuring correct dependency resolution through accurate imports.
    \item \textbf{Constructor Signature.} Captures the instantiation logic of the parameter, providing essential information for creating valid objects.
    \item \textbf{Functional Summary.} Encodes the expected behavioral semantics of the parameter within the function under test, guiding the generation of semantically consistent interactions.
\end{itemize}
This design encapsulates both structural and behavioral perspectives of parameter usage, thereby enriching the contextual signal available to LLM while maintaining a bounded prompt length. 
The combination of these three elements enables the synthesis of test cases that are both executable and semantically aligned with the intended functionality.

Nevertheless, the semantic retrieval phase of type inference may introduce errors, and such erroneous type contexts can inject substantial noise into the downstream test case generation process.
To mitigate this risk, we introduce an \emph{LLM-as-a-Critic} mechanism that evaluates the relevance of candidate type contexts prior to generation.
Instead of inferring parameter types directly, LLM is tasked with validating whether a given type context plausibly corresponds to the parameter at hand.
For example, in the function $login(student)$, if the suggested type context corresponds to a non-entity utility class, the critic model can readily detect the inconsistency.
This validation step substantially improves the robustness of type utilization, thereby enhancing both inference accuracy and the fidelity of generated test cases.

\begin{wrapfigure}{r}{0.5\textwidth}
    \centering 
    \includegraphics[width=0.5\textwidth]{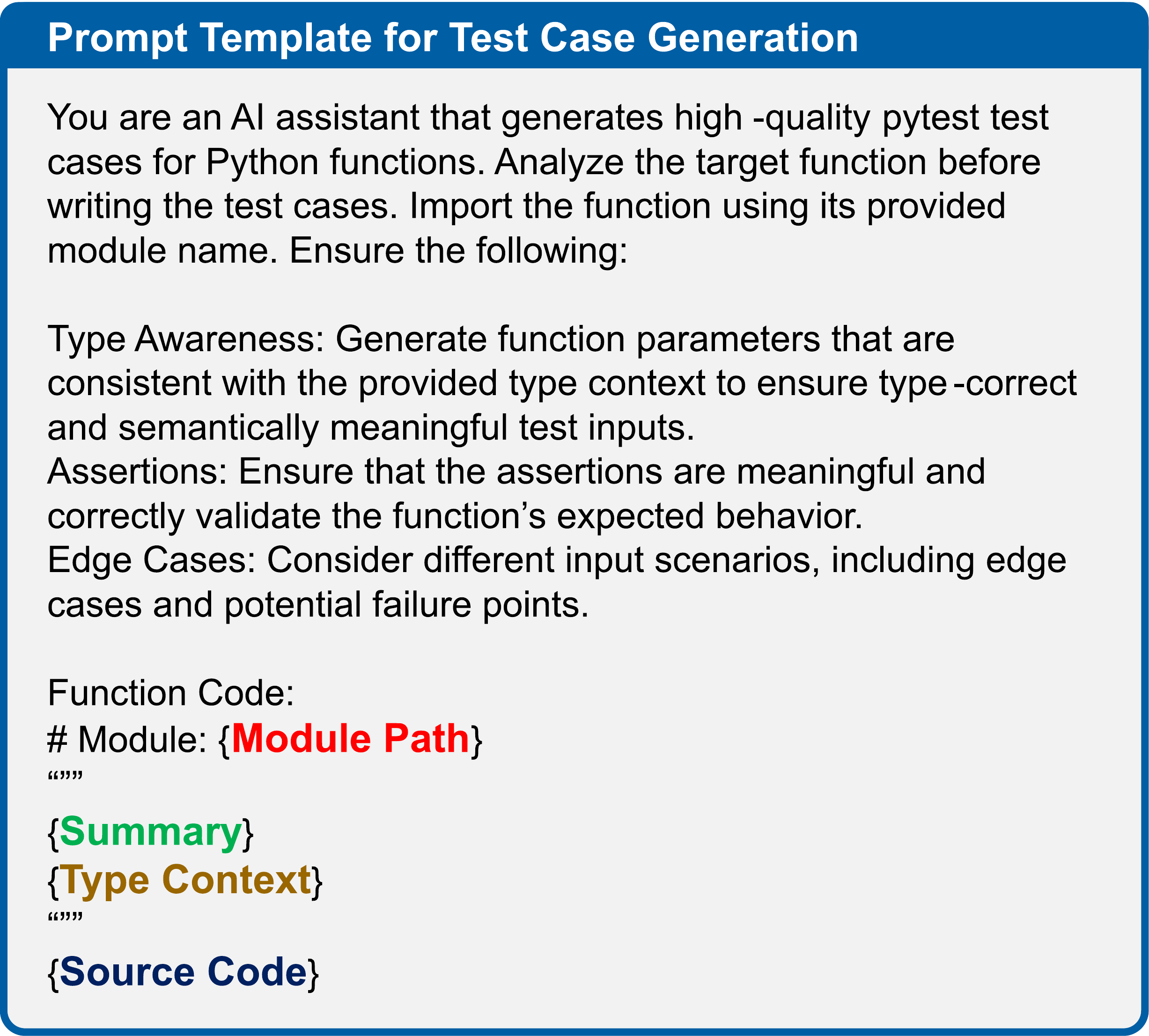}
    \caption{The core part of the prompt for test case generation.}
    \label{fig:prompt} 
\end{wrapfigure}

While the preceding step focuses on parameter-type contexts, we complement it with an overall prompt design that captures broader aspects of test case generation. 
At the system level, we further specify a role-based prompt that explicitly conditions LLM to act as an expert Python developer. 
Empirical studies indicate that role specification improves the consistency of generated code~\cite{white2023prompt}. 
Accordingly, our prompt specifies the target function’s module path to guarantee dependency correctness, enforces \texttt{pytest}-compliant test generation, and incorporates a chain-of-thought prompting strategy to improve the reliability of LLM-based synthesis~\cite{gao2023prompt}. 
Moreover, the prompt integrates the functional summary ($f.\mathit{summary}$) from Section~\ref{subsec:summary} and the parameter type context extracted in the preceding step, guiding the model toward generating test cases that are both semantically accurate and executable.
The final prompt is illustrated in Figure~\ref{fig:prompt}.

\subsubsection{Adaptive Error Repair}

Despite careful prompt engineering, the generated test cases $t$ may still exhibit both syntactic flaws and semantic inconsistencies. 
These errors manifest in diverse forms.
Syntactic errors are primarily manifested as \texttt{ModuleNotFoundError}, which commonly occurs when LLM fails to correctly resolve class dependencies of the function under test or its parameters. 
Semantic errors are more prevalent, with \texttt{AssertionError} and \texttt{AttributeError} being the most frequent. 
\texttt{AssertionError} typically arises from insufficient semantic cues in the prompt or the intrinsic limitations of LLM reasoning, leading to incorrect assertion synthesis. 
\texttt{AttributeError}, in contrast, often arises when an incorrect type leads to invalid member access.

The heterogeneity and unpredictability of these errors render static repair strategies inadequate. 
To address this challenge, we propose an \emph{adaptive error repair mechanism} that exploits LLM’s capacity for context-sensitive reasoning, supplemented with external knowledge retrieval, to iteratively diagnose and rectify faults.
The mechanism proceeds through four stages:
\begin{enumerate}[leftmargin=*]
    \item \textbf{Context-Aware Diagnosis.} LLM analyzes the error trace, localizes the faulty component, and generates candidate rectification strategies. 
    \item \textbf{External Knowledge Augmentation.} Building on the class-level retrieval framework in Section~\ref{subsec:type-infer}, we extend it to the function level. The function summaries are embedded as indexing keys in a vector database, over which LLM-generated queries are executed to retrieve supplementary contextual information when necessary.
    \item \textbf{Repair Synthesis.} Guided by the diagnostic and retrieval results, LLM synthesizes a refined version of the test case that resolves the identified fault. 
    \item \textbf{Iterative Validation and Reduction.} The repaired test case is executed and, if errors persist, the cycle repeats. Once the number of repair attempts exceeds a predefined threshold, a heuristic reduction strategy progressively minimizes the test case until an executable and error-free variant is obtained. 
\end{enumerate}

This adaptive repair strategy equips \toolname with resilience against diverse and unforeseen error patterns, thereby substantially improving the robustness and executability of LLM-generated test cases.

\definecolor{lightyellow}{RGB}{255, 255, 125} 
\section{Evaluation}

In this section, we evaluate \toolname by addressing the following research questions:

\begin{enumerate}[label=\textbf{RQ\arabic*:}, leftmargin=*, align=left]
    \item {How does \toolname compare with state-of-the-art baselines in terms of code coverage?}
    \item {How do different large language models affect the performance of \toolname in test generation?}
    \item {How effective is \toolname’s type inference module compared to existing type inference tools, and what is its impact through ablation analysis?}
    \item {How does incorporating call graph-guided summaries influence the effectiveness of test case generation in \toolname?}
    \item {How does error repair and iterative test case generation improve the quality of test suites produced by \toolname?}
\end{enumerate}

\paragraph{\textbf{Benchmarks}}
We conduct experiments on two complementary benchmarks. The first is the benchmark provided by Pynguin \cite{lukasczyk2022pynguin}, hereafter referred to as \textbf{Pyn}. This benchmark has been widely adopted by tools such as CodaMosa~\cite{Codamosa} and CoverUp~\cite{CoverUp}. 
\textbf{Pyn} consists of 17 real-world Python projects drawn from datasets including BugsInPy~\cite{widyasari2020bugsinpy} and ManyTypes4Py~\cite{mir2021manytypes4py}. 
All modules in this dataset contain type hints compliant with PEP 484\footnote{\url{https://peps.python.org/pep-0484/}}, which defines the standard for type annotations in Python. 
To avoid trivial and untestable modules, we refer to the filtered benchmark provided by CodaMosa~\cite{Codamosa}, which results in a subset of 125 modules.

\begin{table}
\centering
\begin{tabular}{l l l l r}
\toprule
\textbf{Repository} & \textbf{Commit} & \textbf{Modules} & \textbf{Lines} & \textbf{Branches}\\
\midrule
mindsdb/dfsql            & 2c98482  & 10 & 992 & 383 \\
box/genty                & 85f7c96  &  4 &  204 &  72 \\
devshawn/kafka-shell     & 3615895  & 11 &  595 &  170 \\
vitsalis/PyCG            & 8d5dc40  & 20 &  2032 &  961 \\
kieferk/pymssa           & 9d4d3e2  &  3 &  413 &  146 \\
skelsec/pypykatz\_server & bdc76f4  &  5 &  258 &  48 \\
tobgu/pyrthon            & a874549  &  2 &  130 &  33 \\
btwael/superstring.py    & f47ef78  &  1 &  136 &  34 \\
docwza/woa               & 4747379  &  2 &  112 &  24 \\
\bottomrule
\end{tabular}
\caption{Overview of the NA Benchmark: Nine open-source Python projects without type annotations, collected from GitHub.}
\label{tab:na-benchmark}
\end{table}
To complement this type-annotated benchmark, we construct a second dataset, denoted as \textbf{NA}, comprising nine open-source projects with more than 50 GitHub stars but no type annotations. 
We employed AST analysis to select projects in which fewer than 5\% of function parameters contained type annotations.
Furthermore, to mitigate the risk of data leakage, we filtered out projects in which the source code and pre-existing test cases were co-located within the same directory.
As shown in Table~\ref{tab:na-benchmark}, these projects contain 58 modules in total.
Together, \textbf{Pyn} and \textbf{NA} allow us to assess the performance of test generation tools in both type-rich and type-scarce environments.

\paragraph{\textbf{Baselines and Experimental Setup}}
We compare \toolname against two state-of-the-art LLM-based unit test generation tools: CodaMosa~\cite{Codamosa} and CoverUp~\cite{CoverUp}.  
CodaMosa integrates large language models with search-based testing, iteratively generating new test cases to overcome coverage bottlenecks.  
CoverUp adopts a complementary strategy by combining LLMs with coverage-guided feedback to generate high-coverage regression tests.

To ensure fair evaluation, we align experimental configurations across tools:  
\begin{itemize}[leftmargin=*]
\item For \toolname, we adopt a retrieval-augmented generation configuration, using the "BAAI/bge-large-en-v1.5" model for embeddings\footnote{https://huggingface.co/BAAI/bge-large-en-v1.5}, and Chroma\footnote{https://github.com/chroma-core/chroma} for building behavior-based indices. 
\item We employ \texttt{gpt-4o} as the underlying LLM for all tools by default, with consistent model parameters. \toolname and CoverUp are configured to perform three rounds of test generation, producing one test per function in each round. 
\item For CodaMosa, we follow the original protocol, executing three rounds of testing with a 10-minute budget per module. 
Since the original CodaMosa employed Codex~\cite{chen2021evaluating}, which produces code-completion style outputs, we adapted its output parsing to align with conversational LLMs, ensuring comparability across tools.
\end{itemize}

\subsection{RQ1: Effectiveness of \toolname}

This research question investigates the effectiveness of \toolname in automated test generation. 
Specifically, we examine whether \toolname achieves higher test coverage than existing state-of-the-art approaches and whether it maintains stable performance across different benchmarks. 
To this end, we conduct quantitative comparisons with CoverUp and CodaMosa using standard coverage metrics and further complement the analysis with a qualitative case study.

\begin{figure}[htbp]
\centering
\begin{minipage}{0.49\textwidth}
\centering
\includegraphics[width=\textwidth]{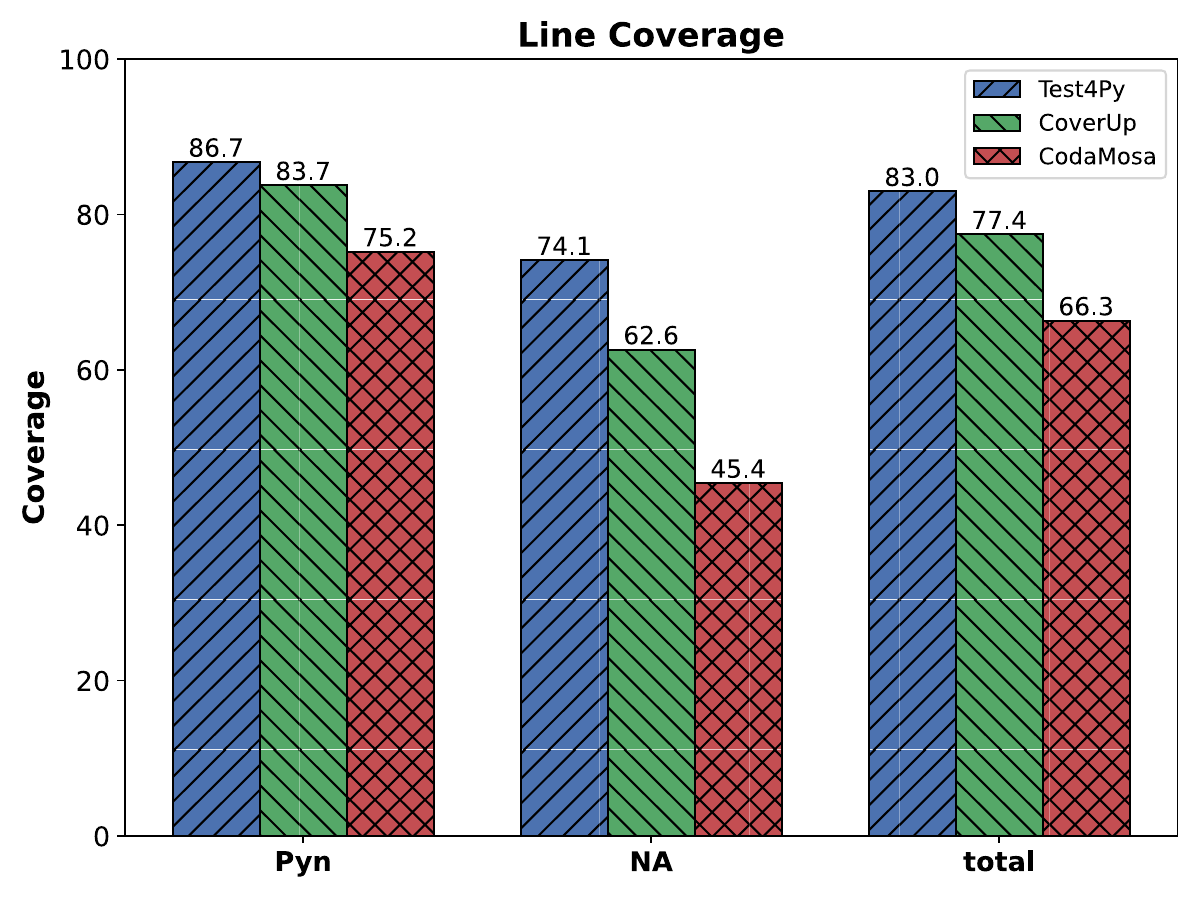}
\caption{Line Coverage Comparison}
\label{fig:Line_coverage}
\end{minipage}
\hfill
\begin{minipage}{0.49\textwidth}
\centering
\includegraphics[width=\textwidth]{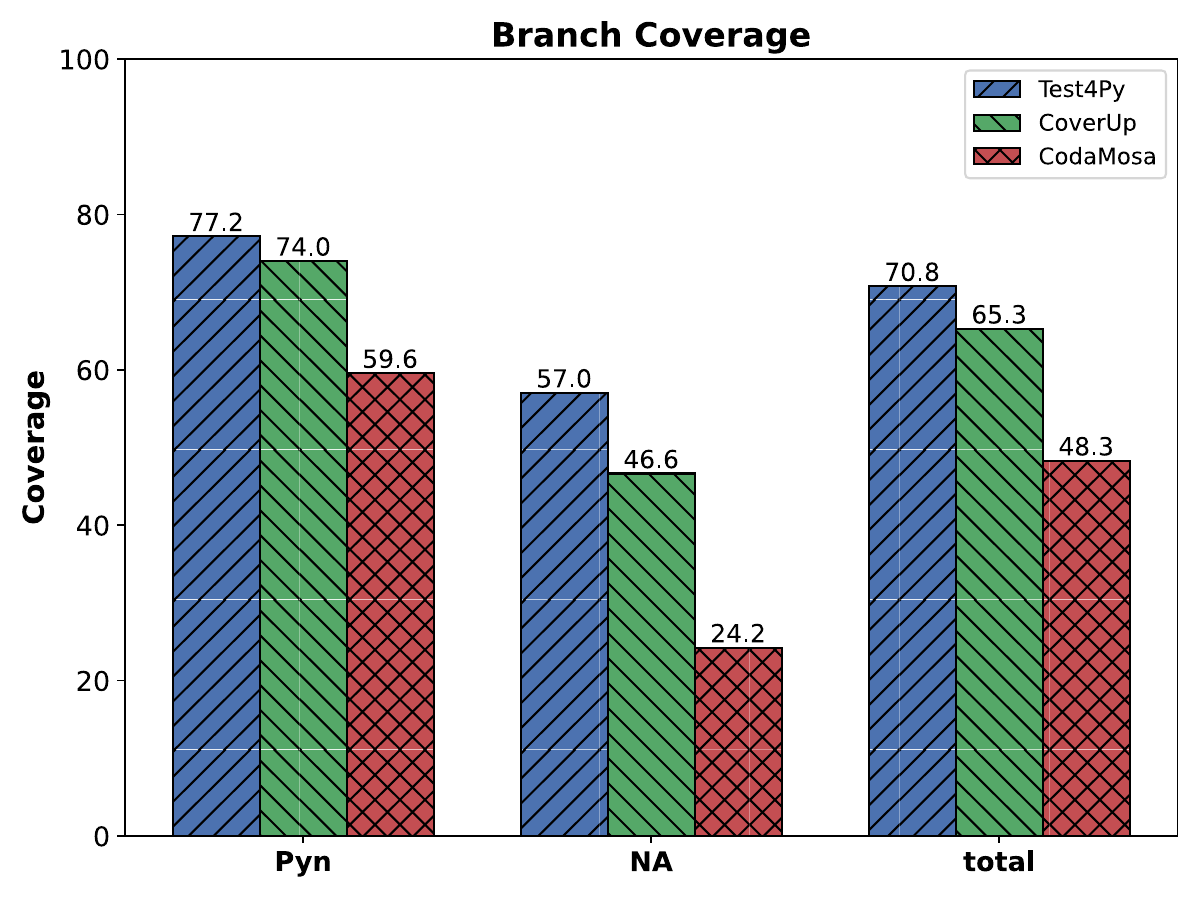}
\caption{Branch Coverage Comparison}
\label{fig:Branch_coverage}
\end{minipage}
\label{fig:line-branch}
\end{figure}

\paragraph{\textbf{Test Coverage.}}
Following the evaluation methodology of CoverUp, 
we measure both line coverage and branch coverage across benchmarks. 
Figures \ref{fig:Line_coverage} and \ref{fig:Branch_coverage} report the results. 
On average, \toolname attained the highest line coverage (83.0\%), surpassing CoverUp (77.4\%) and CodaMosa (66.3\%). 
Similarly, it achieved the highest branch coverage (70.8\%), outperforming CoverUp (65.3\%) and CodaMosa (48.3\%). 
These results demonstrate that \toolname consistently generates test suites that achieve broader exploration of program behaviors.

A closer inspection reveals different patterns across datasets. 
On the \textbf{Pyn} benchmark, the gap between \toolname and CoverUp was relatively small. 
This is largely attributed to CoverUp’s mechanism of dynamically retrieving function or class definitions through function calls, which is particularly effective in \textbf{Pyn} where most parameters are annotated with type hints. 
However, in the \textbf{NA} benchmark, which lacks type hints, CoverUp’s mechanism becomes less effective, leading to a substantial performance decline. 
Consequently, in \textbf{NA}, \toolname outperformed CoverUp by 18.4 percentage in line coverage and 22.3 in branch coverage, compared to smaller differences of 7.2 and 8.4 when averaged across both benchmarks.

\begin{figure}[htbp]
\centering
\begin{minipage}{0.49\textwidth}
\centering
\includegraphics[width=\textwidth]{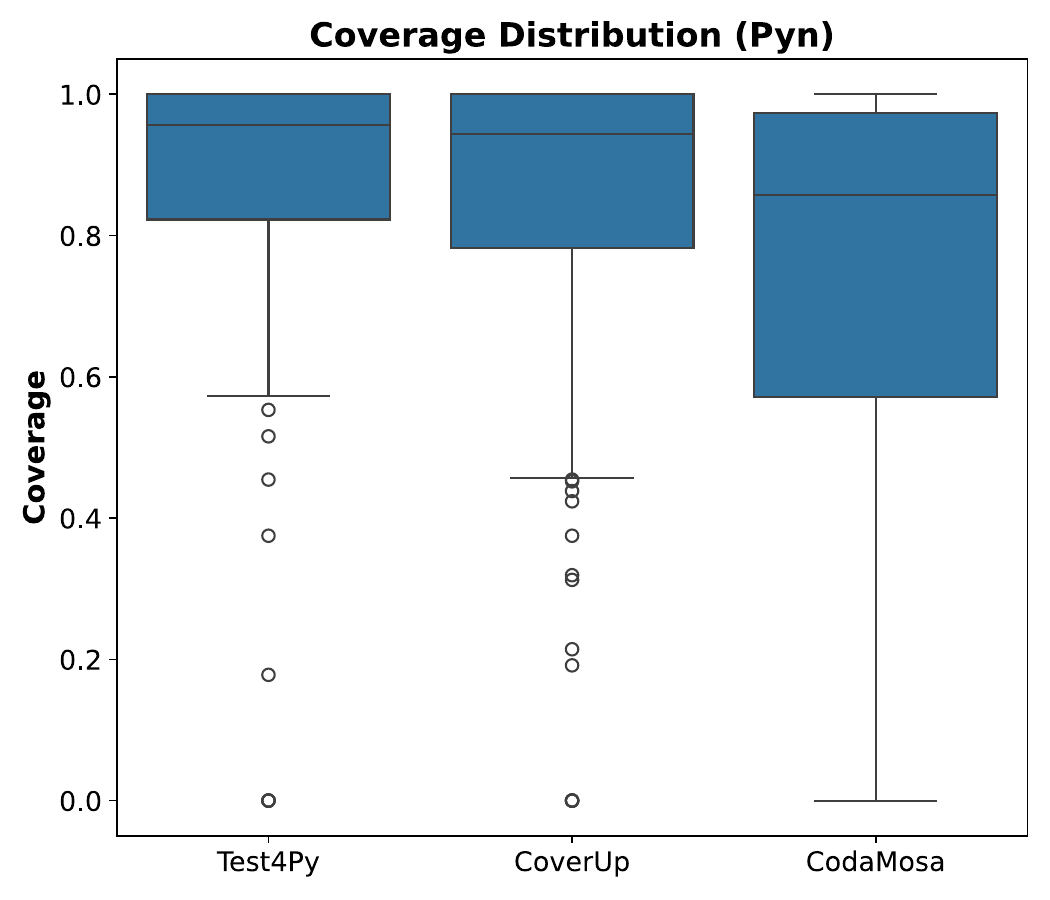}
\caption{Coverage Distribution on Pyn Dataset}
\label{fig:Pyn_coverage}
\end{minipage}
\hfill
\begin{minipage}{0.49\textwidth}
\centering
\includegraphics[width=\textwidth]{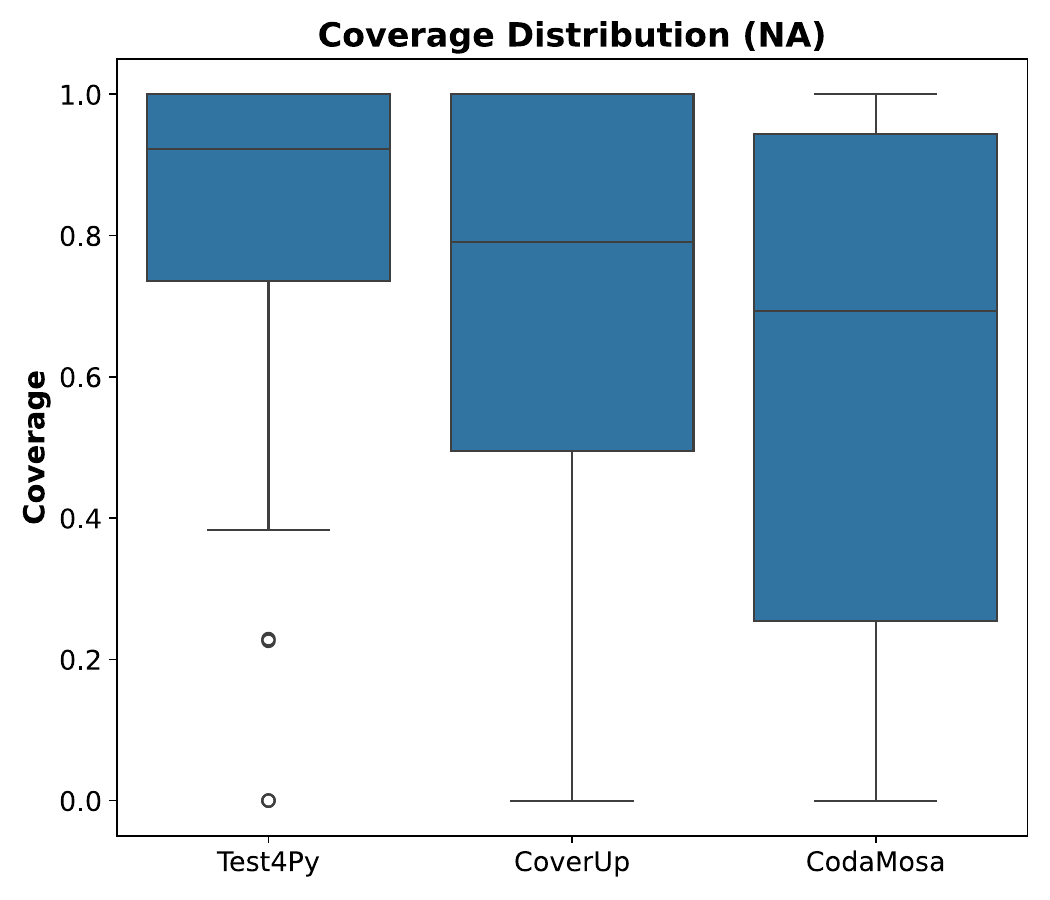}
\caption{Coverage Distribution on NA Dataset}
\label{fig:NA_coverage}
\end{minipage}
\end{figure}

\paragraph{\textbf{Stability of Coverage.}}
To evaluate performance stability, we analyzed the distribution of the line coverage at the module level
using box plots, as shown in Figure \ref{fig:Pyn_coverage} and \ref{fig:NA_coverage}.
The median (Q2) indicates the central trend, while the interquartile range (IQR) reflects the variability between modules. 
A smaller IQR indicates more stable performance.

On the \textbf{Pyn} benchmark, \toolname achieved a median coverage of 95.7, slightly higher than CoverUp (94.4).
More importantly, its IQR (17.7) was considerably lower than that of CoverUp (21.7) and CodaMosa (40.2), reflecting its robustness across diverse modules. 
On the \textbf{DT} benchmark, \toolname exhibited even stronger advantages, with a median coverage of 92.2, exceeding CoverUp (79.1) and CodaMosa (69.3). 
Its IQR (26.5) was the smallest among all tools, representing only 52.5\% of CoverUp’s and 38.5\% of CodaMosa’s. 
These results confirm that \toolname achieves not only higher coverage but also more consistent performance across modules.

\paragraph{\textbf{Case Study.}}
To gain qualitative insights, we analyzed 15 modules from the \textbf{NA} benchmark in which \toolname exhibited the largest improvements over CoverUp. Our manual inspection identifies two recurring factors that account for CoverUp’s limited performance:
\begin{itemize}[leftmargin=*]
\item \textbf{Incorrect construction of function parameter types} (7 cases). 
These errors led to test generation failures, such as member access exceptions caused by type mismatches.
\item \textbf{High-level functions with extensive external dependencies} (4 cases). 
Testing was hindered by complex call structures, including functions with numerous nested invocations or those that merely acted as wrappers delegating execution to other components.
\end{itemize}
These findings underscore the importance of precise type information and demonstrate that \toolname’s refinement mechanisms are effective in mitigating such challenges.

\begin{tcolorbox}
\textbf{Summary:} \toolname consistently achieves higher and more stable coverage than state-of-the-art baselines. Its advantages are especially pronounced in scenarios without type hints, underscoring its effectiveness in dynamically typed settings.
\end{tcolorbox}

\subsection{RQ2: The Impact of Different Large Language Models on \toolname}
\label{subsec:diif_models}

This research question examines how the choice of large language model (LLM) influences the effectiveness of \toolname. 
In particular, we aim to assess (i) whether different LLMs exhibit varying levels of test generation capability, and (ii) whether the observed differences may be attributed to potential data contamination in the training corpora.

\begin{figure}[htbp]
\centering
\begin{minipage}{0.49\textwidth}
\centering
\includegraphics[width=\textwidth]{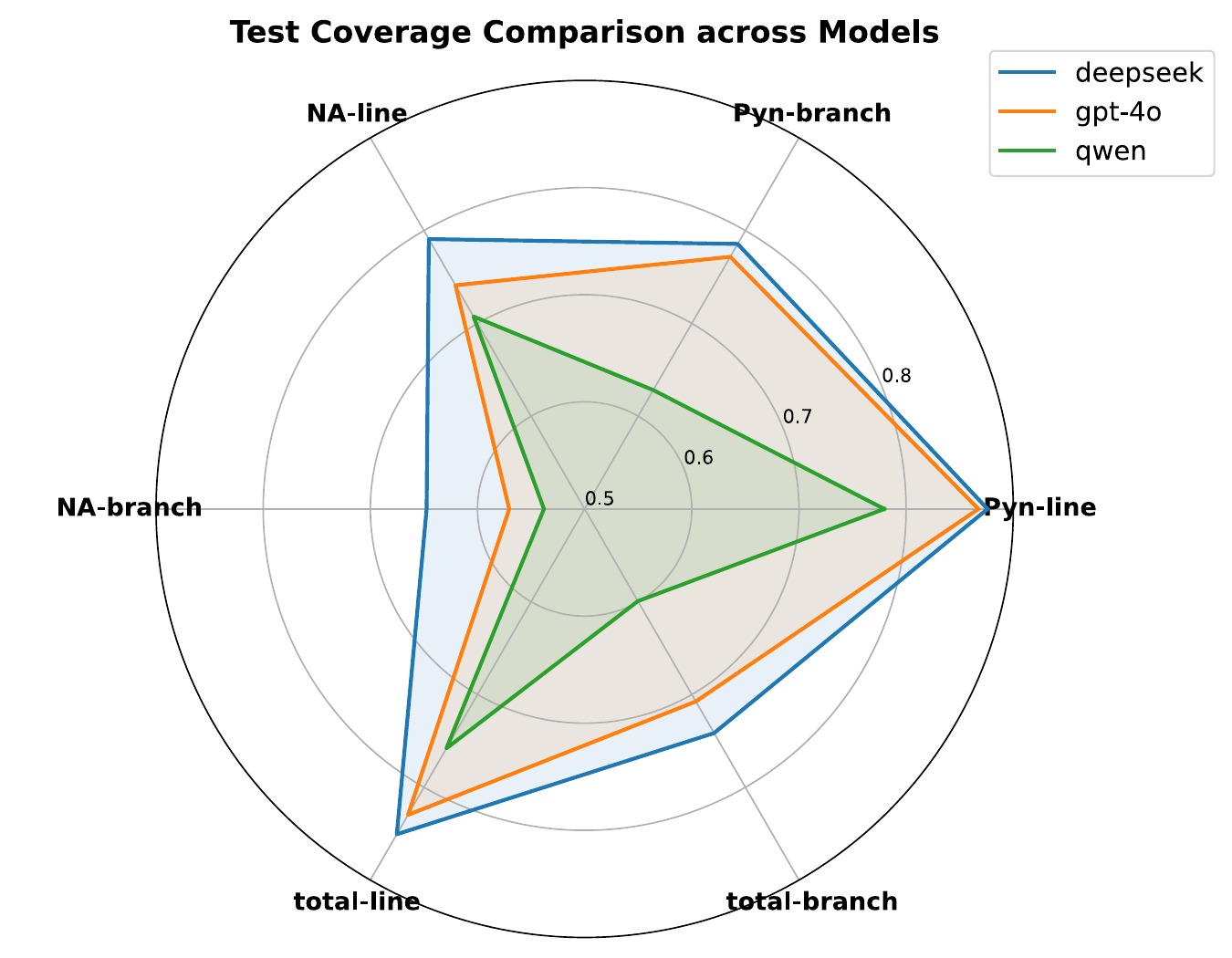}
\caption{Line and Branch Coverage of Different LLMs across Benchmarks}
\label{fig:model_coverage}
\end{minipage}
\hfill
\begin{minipage}{0.49\textwidth}
\centering
\includegraphics[width=\textwidth]{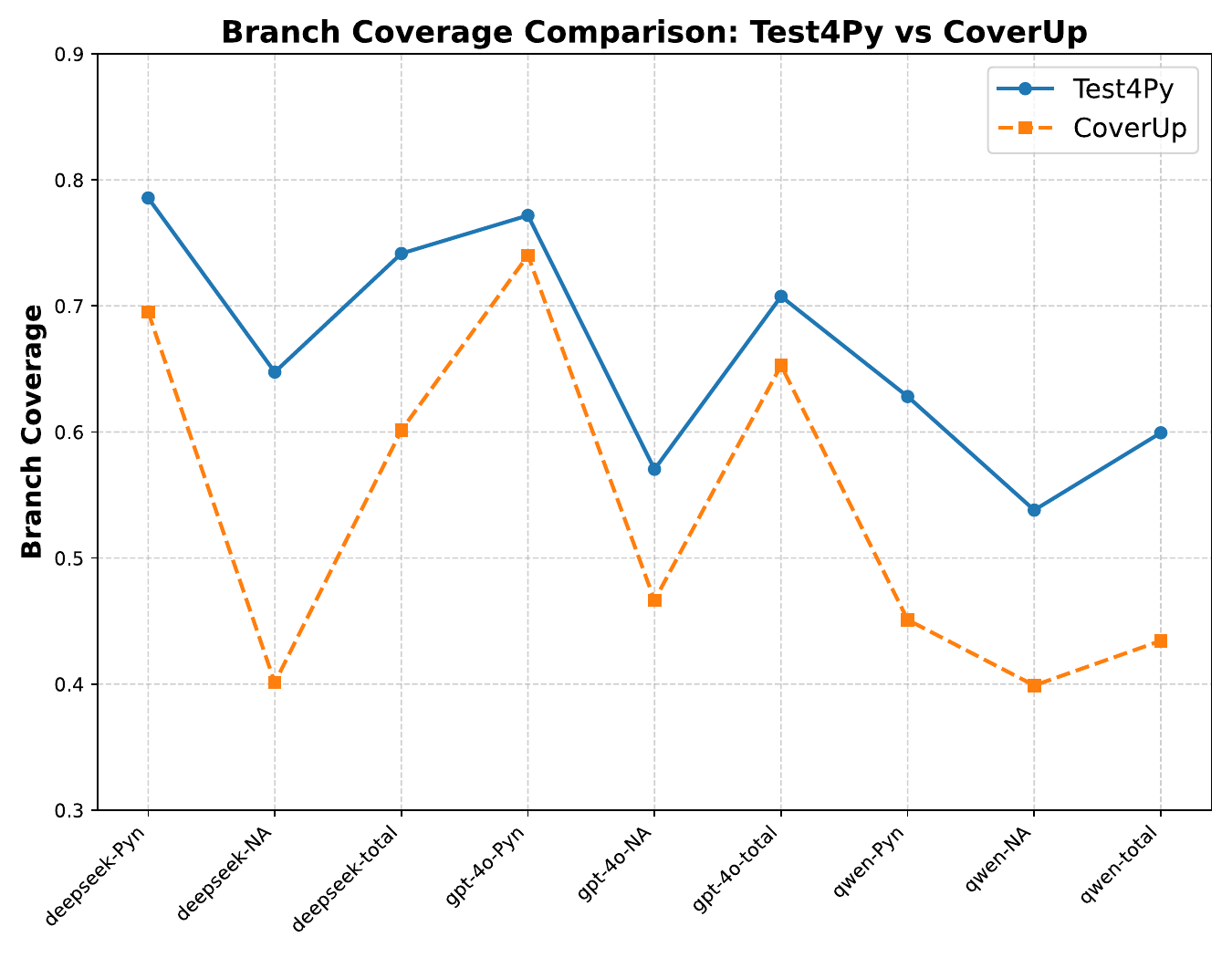}
\caption{Comparison of Branch Coverage between \toolname and CoverUp across Different Models}
\label{fig:coverup_models}
\end{minipage}
\end{figure}

We evaluated \toolname using three LLMs: \texttt{gpt-4o-2024-05-13}, \texttt{deepseek-v3-250324}, and \texttt{qwen-plus} (hereafter referred to as \textit{gpt-4o}, \textit{deepseek}, and \textit{qwen}, respectively). 
The models were tested on both  \textbf{Pyn} and \textbf{NA}, and their performance was measured in terms of line and branch coverage. 
Figure~\ref{fig:model_coverage} presents the results in a radar chart, with six dimensions corresponding to line and branch coverage across the two datasets.

Overall, \textit{deepseek} achieved the highest performance, with an average line coverage of 85.1\% and branch coverage of 74.2\%, consistently outperforming the other models across all dimensions. 
\textit{gpt-4o} ranked second, showing competitive performance that was close to \textit{deepseek} on the \textbf{Pyn} benchmark but lagging more substantially on the \textbf{NA} benchmark. 
This suggests that more complex type inference tasks may benefit from more precise prompt engineering. 
\textit{Qwen} exhibited the lowest overall coverage, though its results were relatively stable across the two datasets.

We further investigate the impact of different underlying LLMs on CoverUp, with the results presented in Figure~\ref{fig:coverup_models}. 
Across all models and benchmarks, \toolname consistently outperforms CoverUp. 
CoverUp achieves its highest performance with \textit{gpt-4o}, but its branch coverage decreases by 7.9\% when switching to \textit{deepseek}. 
In contrast, \toolname exhibits a 4.8\% performance gain when using \textit{deepseek} compared to \textit{gpt-4o}. 
The performance gap is even more pronounced for \textit{qwen}: CoverUp suffers a 33.5\% drop relative to \textit{gpt-4o}, while \toolname only declines by 15.3\%.  
A deeper analysis suggests that this discrepancy may stem from CoverUp’s reliance on function calls, making it more sensitive to prompt engineering and the intrinsic capabilities of the underlying models. 
In comparison, \toolname demonstrates stronger cross-model generalization, maintaining robust performance even when the model quality varies.

Given that LLMs are typically trained on large-scale corpora of open-source code, there exists a risk of data contamination, whereby evaluation benchmarks may inadvertently overlap with training data. 
Such contamination could lead to the reproduction of pre-existing test cases rather than the generation of novel ones. 
To further assess this risk, we selected projects with pre-existing test suites and compared the generated test cases against the original ones using \texttt{pycode\_similar}\footnote{\url{https://github.com/fyrestone/pycode_similar}}, an AST-based similarity detection tool.  

The results indicate that the average similarity scores for \toolname were 0.712 with \textit{deepseek}, 0.670 with \textit{gpt-4o}, and 0.682 with \textit{qwen}. In contrast, CoverUp exhibited even higher similarity scores of 0.716, 0.725, and 0.731 under the same models, respectively. 
These findings suggest that \toolname is less prone to reusing existing test cases compared to CoverUp.

\begin{tcolorbox}
\textbf{Summary:} Different LLMs have a measurable impact on the performance of \toolname. 
\textit{Deepseek} demonstrates the highest coverage, followed by \textit{gpt-4o} and \textit{qwen}. 
We further evaluated state-of-the-art baselines across the same models and observed that \toolname exhibits greater stability across model variations, highlighting its stronger robustness to differences in underlying LLM capabilities.
\end{tcolorbox}

\subsection{RQ3: The Effectiveness and Ablation of \toolname's Type Inference}

This research question evaluates the performance of \toolname's type inference process compared to state-of-the-art (SOTA) tools and presents an ablation study of the type inference component.

Numerous Python type-inference tools have been developed, such as Type4Py \cite{mir2022type4py} and Hityper \cite{peng2022static}. 
Among these, Hityper effectively integrates static inference with deep learning techniques. 
Recently, with the advancement of LLMs, their type inference performance has even surpassed that of Hityper \cite{peng2023generative}. 
Therefore, we chose Type4Py, Hityper and an LLM-Only baseline for our comparison.
To further investigate the individual contributions of syntactic filtering and semantic retrieval in our framework, we conducted ablation studies. 
Specifically, we denote the variant that applies only syntactic filtering as \toolname-sy, and the variant that employs only semantic retrieval as \toolname-se.

\begin{wrapfigure}{r}{0.5\textwidth}
    \centering 
    \includegraphics[width=0.5\textwidth]{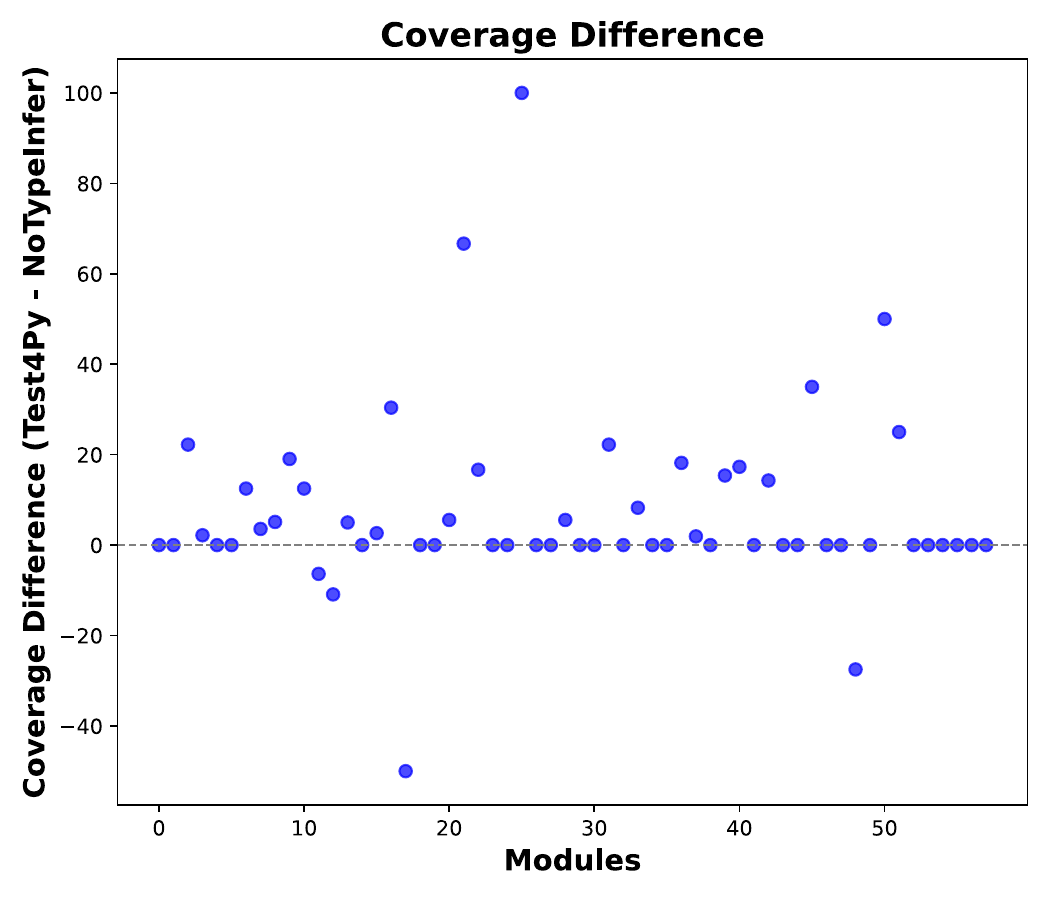}
    \caption{Coverage difference with and without \toolname's type inference module}
    \label{fig:type_variant} 
\end{wrapfigure}

The \textbf{Pyn} benchmark contains a substantial number of type annotations. 
We removed these function parameter annotations and treated them as the ground truth, thus creating the \textbf{Pyn-type} benchmark. 
This benchmark includes 222 User-defined types and 1806 Non-user-defined types. We define an Exact Match as a strict match and a Relaxed Match as a loose match (e.g., for List[int], only the outermost List type is checked). 
We ran the six aforementioned tools on the \textbf{Pyn-type} benchmark, and the results are presented in Table~\ref{tab:type_prediction_results}.

\begin{table}[t!]
\centering
\begin{tabular}{llrrrr}
\toprule
\multirow{2}{*}{Setting} & \multirow{2}{*}{Type Category} & \multicolumn{2}{c}{Relaxed Match} & \multicolumn{2}{c}{Exact Match} \\
\cmidrule(lr){3-4} \cmidrule(lr){5-6}
 & & Success & Failure & Success & Failure \\
\midrule
\multirow{2}{*}{\textsc{\toolname}} 
 & User-defined      & \textbf{141 (63.5\%)} & 81 (36.5\%)  & \textbf{81 (36.5\%)}  & 141 (63.5\%) \\
 & Non-user-defined  & 1230 (68.1\%) & 576 (31.9\%) & \textbf{1006 (55.7\%)} & 800 (44.3\%) \\
 \midrule
\multirow{2}{*}{\textsc{\toolname-se}} 
 & User-defined      & 137 (61.7\%) & 85 (38.3\%)  & 80 (36.0\%)  & 142 (64.0\%) \\
 & Non-user-defined  & \textbf{1238 (68.5\%)} & 568 (31.5\%) & 1001 (55.4\%) & 805 (44.6\%) \\
 \midrule
\multirow{2}{*}{\textsc{\toolname-sy}} 
 & User-defined      & 117 (52.7\%) & 105 (47.3\%)  & 67 (30.2\%)  & 155 (69.8\%) \\
 & Non-user-defined  & 1222 (67.7\%) & 584 (32.3\%) & 992 (54.9\%) & 814 (45.1\%) \\
\midrule
\multirow{2}{*}{\textsc{LLM-Only}} 
 & User-defined      & 99 (44.6\%)  & 123 (55.4\%) & 53 (23.9\%)  & 169 (76.1\%)\\
 & Non-user-defined  & 1221 (67.6\%) & 585 (32.4\%) & 986 (54.6\%) & 820 (45.4\%)\\
\midrule
\multirow{2}{*}{{Type4Py}} 
 & User-defined      & 31 (14.0\%)  & 191 (86.0\%) & 23 (10.4\%)  & 199 (89.6\%)\\
 & Non-user-defined  & 852 (47.2\%) & 954 (52.8\%) & 751 (41.6\%) & 1055 (58.4\%)\\
\midrule
\multirow{2}{*}{{HiTyper}} 
 & User-defined      & 1 (0.5\%)   & 221 (99.5\%) & 0 (0\%)   & 222 (100\%) \\
 & Non-user-defined  & 259 (14.3\%) & 1547 (85.7\%) & 246 (13.6\%) & 1560 (86.4\%)\\
\bottomrule
\end{tabular}
\caption{Type Prediction Results: Relaxed vs. Exact Matching, split by User-defined and Non-user-defined types.}
\label{tab:type_prediction_results}
\end{table}

As shown in the results, \toolname achieves the highest prediction performance among all evaluated approaches. Its accuracy on user-defined types reaches 63.5\%, substantially outperforming the baseline LLM-Only, which reaches 44.6\%.
For non-user-defined types, \toolname achieves an accuracy of 68.1\%, which is comparable to LLM-Only’s 67.6\%. 
In contrast, Type4Py demonstrates particularly weak performance on user-defined types, suggesting that traditional machine learning-based approaches are not well-suited for this task. 
HiTyper exhibits consistently poor performance across all categories, primarily due to its inability to handle certain language constructs (e.g., ``for loops with else statements’’), which prevents it from producing valid results. 
These findings highlight the limitations of conventional type inference tools and underscore the robustness and effectiveness of \toolname’s inference mechanism.

The precision of \toolname-se is comparable to that of \toolname, indicating that the semantic retrieval mechanism achieves high precision, accurately identifying the correct type in most cases even without syntactic filtering. 
However, this does not imply that syntactic filtering is ineffective.
In contrast, it significantly reduces the computational cost of retrieval. 
Specifically, when syntactic filtering is disabled, each query produces an average of 41.38 candidate classes; incorporating syntactic filtering reduces this number to 28.74, achieving a reduction of 30.5\%. 
Although \toolname-sy exhibits a noticeable performance drop compared to \toolname, it still substantially outperforms the LLM-Only baseline, further confirming the effectiveness of syntactic filtering. 

To further investigate the contribution of \toolname’s type inference to test case generation, we conducted an ablation study. 
Specifically, we implemented a variant, \textit{NoTypeInfer}, which disables the type inference component. 
Both versions were evaluated on the \textbf{NA} benchmark, and the differences in branch coverage are illustrated in Figure~\ref{fig:type_variant}. 
Among the 58 modules, \toolname achieved higher coverage in 25 modules and similar the coverage of \textit{NoTypeInfer} in 29 modules. 
We conducted a detailed analysis of the remaining four modules where coverage decreased.
Three of these degradations can be attributed to the inherent non-determinism of LLM outputs, while one case was likely due to erroneous type inference introducing misleading contextual cues into the prompt. 
Although the type inference process still exhibits a relatively high error rate, \toolname mitigates its negative effects by leveraging the LLM’s capacity to filter irrelevant information, thereby limiting adverse impact to a small minority of cases. 
Overall, these findings demonstrate that type inference consistently enhances the effectiveness of \toolname in scenarios where type annotations are absent.

\begin{tcolorbox}
\textbf{Summary:} \toolname's type inference is more effective than Hityper and LLM-Only, especially when handling user-defined types. 
Additionally, the type inference system enables \toolname to achieve better performance in the absence of type annotations.
\end{tcolorbox}

\subsection{RQ4: The Role of Call Graph-Based Summaries}

\begin{wrapfigure}{r}{0.5\textwidth}
    \centering 
    \includegraphics[width=0.5\textwidth]{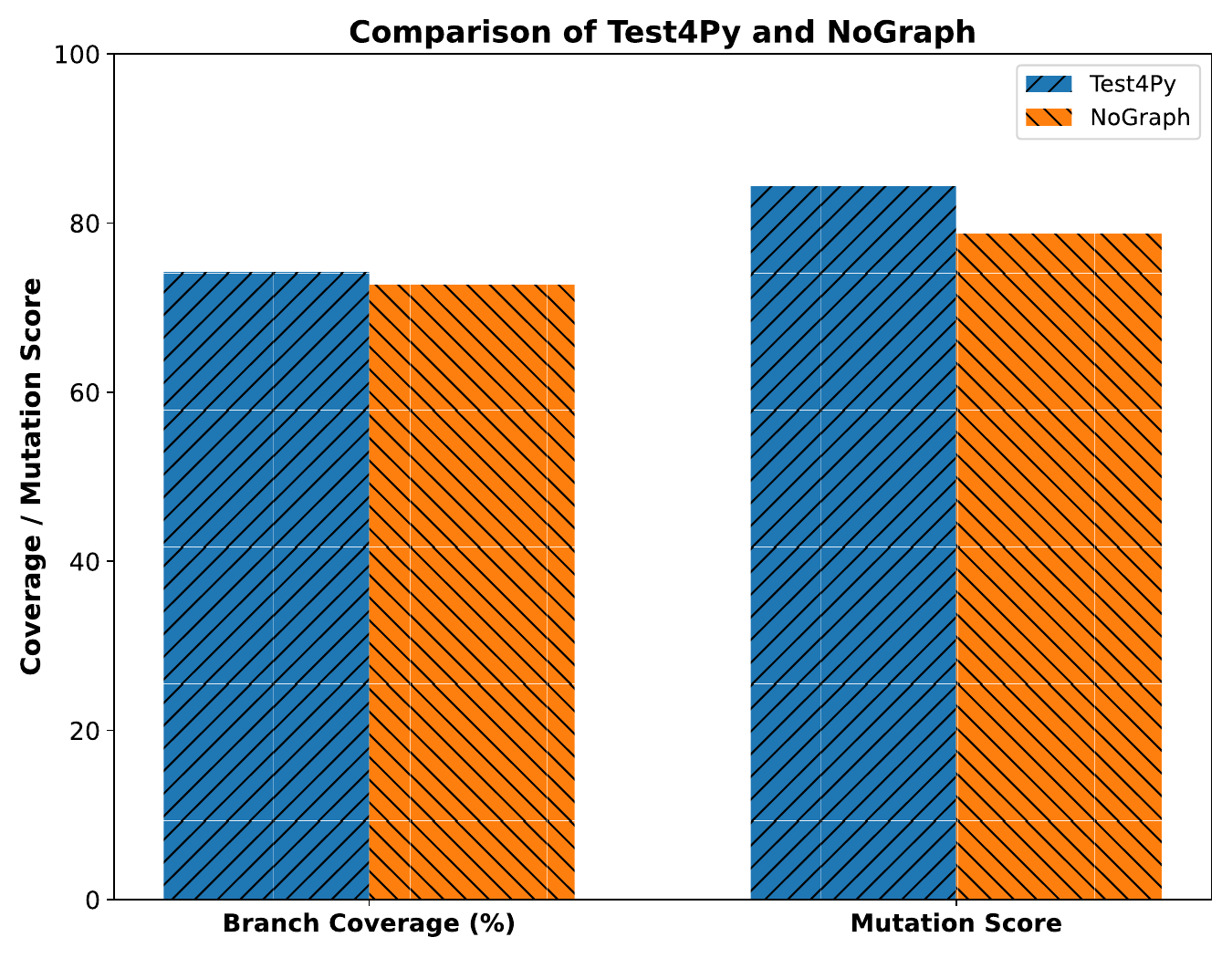}
    \caption{Mutation score and branch coverage of \toolname and NoGraph}
    \label{fig:mutation_score} 
\end{wrapfigure}

This research question investigates the effectiveness of incorporating call graph-guided information into function summaries for test case generation. 
Specifically, we evaluate whether interprocedural context captured by call graph analysis improves the semantic adequacy and fault-detection capability of the generated test cases. 
To this end, we conducted an ablation study by creating a variant, \textsc{NoGraph}, in which the prompts exclude any contextual information derived from call graph analysis.

We first compared \toolname and \textsc{NoGraph} on two benchmarks in terms of branch coverage.
\toolname achieved 74.2\% coverage, while \textsc{NoGraph} reached 72.6\%. 
Although the improvement appears modest, the benefit of call graph-guided summaries extends beyond raw coverage. 
By providing interprocedural semantic context, these summaries guide the LLM to generate more behaviorally rich test cases. 
For example, in the $dataclasses\_json.utils$ module, \textsc{NoGraph} primarily produced trivial input-output identity checks, whereas \toolname generated test cases that captured more realistic usage scenarios of the function under test.

Since coverage is not always strongly correlated with fault-detection capability, we further assessed the effectiveness of call graph-guided summaries using mutation testing. 
Mutation testing introduces syntactically valid code transformations (mutations) and evaluates whether the generated test suite can detect them, providing a more reliable measure of regression testing strength than coverage alone \cite{zhang2015assertions}. 
However, applying mutation testing to compare \toolname with existing tools such as CoverUp and CodaMosa would be unfair, as these tools explicitly optimize for minimal test suites without sacrificing coverage. 
Moreover, mutation testing incurs significantly higher computational overhead. 
For these reasons, we restricted our comparison to \toolname and \textsc{NoGraph}.

We employed \texttt{mutmut}\footnote{\url{https://github.com/boxed/mutmut}} to compute mutation scores. 
After excluding projects where \texttt{mutmut} could not be executed, the results are reported in Figure~\ref{fig:mutation_score}. 
On average, \toolname achieved a mutation score of 0.843, compared to 0.787 for \textsc{NoGraph}. 
This consistent improvement across multiple projects quantitatively demonstrates the contribution of call graph-guided summaries to fault-detection effectiveness.


\begin{tcolorbox}
\textbf{Summary:} Call graph-guided summaries improve both branch coverage and mutation score, thereby enhancing the semantic adequacy and fault-detection capability of the generated test cases. 
\end{tcolorbox}

\subsection{RQ5: Test Case Repair and Iterative Generation}

This research question examines the role of automated repair and the effectiveness of iterative test case generation in improving the overall quality of the generated test suite. 
Specifically, we analyze (i) the types of errors encountered and the effectiveness of the repair mechanism, and (ii) the distribution of execution time across different stages of test case generation.

\begin{figure}[htbp]
\centering
\begin{minipage}{0.49\textwidth}
\centering
\includegraphics[width=\textwidth]{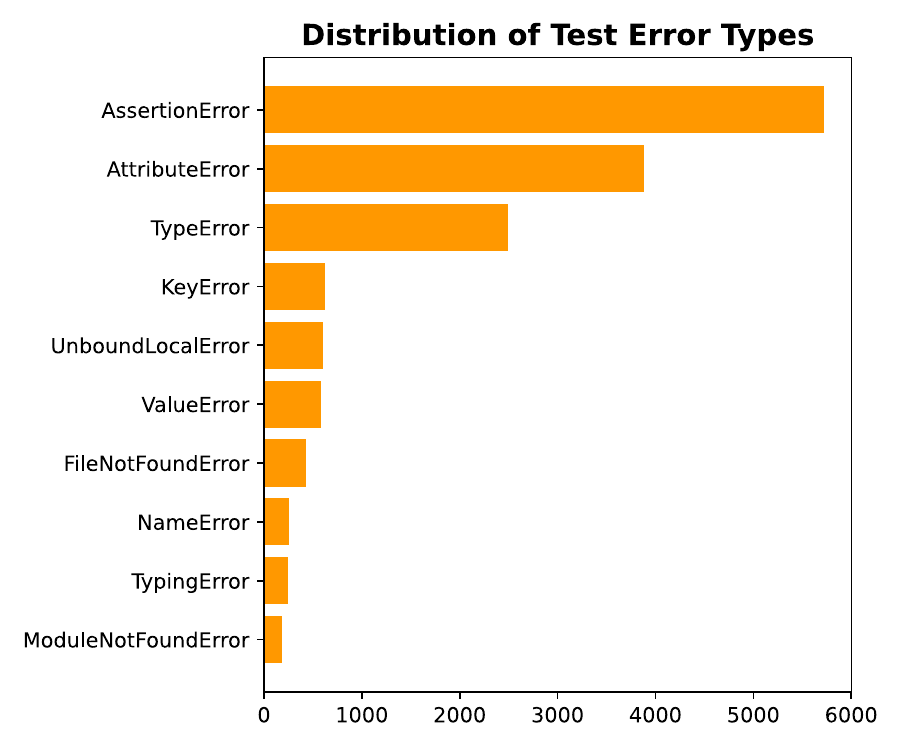}
\caption{Distribution of the ten most frequent error types.}
\label{fig:test_error_types_distribution}
\end{minipage}
\hfill
\begin{minipage}{0.49\textwidth}
\centering
\includegraphics[width=\textwidth]{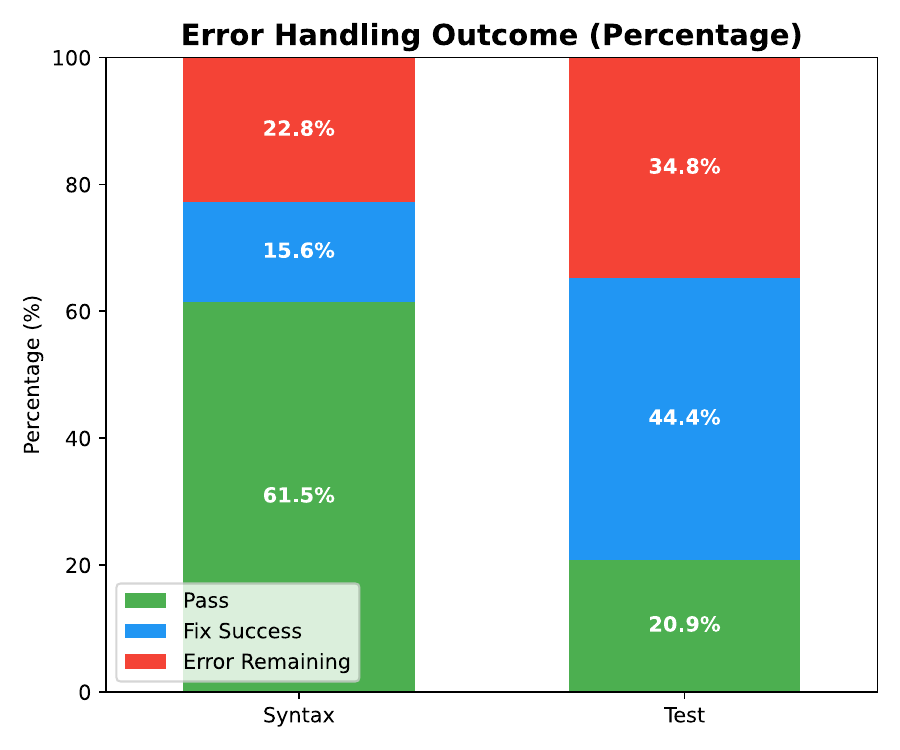}
\caption{Error handling outcomes for syntax and test errors.}
\label{fig:error_handling_outcome}
\end{minipage}
\end{figure}

\paragraph{\textbf{Error Characterization and Repair Effectiveness}}
We analyzed the top-10 error types observed in the generated tests (Figure~\ref{fig:test_error_types_distribution}). 
The most frequent error is \texttt{AssertionError}, highlighting the difficulty LLMs face in generating precise and semantically valid assertions~\cite{zhang2025exploring}. 
The next most common errors, \texttt{AttributeError} and \texttt{TypeError}, indicate persistent challenges related to type inference and type consistency. 
Despite recent progress in type-aware generation, type-related issues remain a key bottleneck.

Repair outcomes are summarized in Figure~\ref{fig:error_handling_outcome}. 
For syntax errors, the initial correctness rate was 61.5\%, and among the remaining erroneous tests, 40.6\% were successfully repaired.
For semantic test errors, only 20.9\% of the tests passed initially, but the repair mechanism successfully resolved 56.1\% of the failing tests.
The lower initial pass rate is largely attributable to LLM’s strategy of generating multiple assertions per test, which decreases the probability of immediate success. 
Nevertheless, iterative repair effectively transforms failing test cases into passing ones, demonstrating the robustness of the repair module.

\paragraph{\textbf{Execution Time and Coverage Evolution}}
Table~\ref{table:time} reports the relative time cost of summarization and each generation phase, together with the coverage achieved. 
The first generation dominates execution time since \toolname must generate tests for all uncovered functions. 
In subsequent iterations, functions that have already achieved full coverage are excluded, resulting in progressively shorter generation phases. 
Importantly, coverage consistently improves across iterations (from 65.2\% to 74.1\%), suggesting that iterative generation can substantially increase the achievable coverage. 
Removing the limit on the number of iterations may therefore yield even higher coverage.

\begin{table}[htbp]
\centering
\begin{tabular}{lcc}
\toprule
\textbf{Stage} & \textbf{Time Proportion} & \textbf{Coverage Achieved} \\
\midrule
Summarization & 27.6\% & -- \\
Generation 1 & 40.4\% & 0.652 \\
Generation 2 & 17.5\% & 0.712 \\
Generation 3 & 14.5\% & 0.741 \\
\bottomrule
\end{tabular}
\caption{Execution time distribution and coverage improvement across generations.}
\label{table:time}
\end{table}

CodaMosa takes 1,485 seconds to generate test cases per module.
\toolname requires 690 seconds per module, which is slower than CoverUp's 275 seconds. 
Considering that test case generation is typically an offline process within the software development lifecycle, the additional execution time represents a reasonable trade-off for higher coverage and test suite quality.

\begin{tcolorbox}
\textbf{Summary:} Automated repair substantially mitigates both syntax- and semantics-related errors, ensuring that the final test suite is both comprehensive and reliable.
Although \toolname incurs a higher execution cost, the resulting increase in coverage and suite quality justifies the additional time. 
\end{tcolorbox}

\section{Threats to Validity}
While \toolname does not directly replicate existing test cases, the broader challenge of data contamination remains a fundamental concern for the reliability of evaluating LLM-based testing frameworks.
Our empirical analysis of test cases generated by CoverUp uncovers instances where the model exhibited knowledge of task-relevant information despite the absence of explicit contextual input, providing evidence of potential leakage from pretraining corpora.
To mitigate this risk, future work will focus on developing a continually updated benchmark dataset that minimizes contamination and ensures the robustness and fairness of empirical evaluations.
Moreover, \toolname currently lacks specialized handling for less prevalent third-party libraries, those that do not appear in the target project and are unlikely to be represented in the LLM’s training data.
This limitation highlights a promising direction for further improvement, and future research will extend \toolname to better support such underrepresented libraries.

\section{Related Work}

Our work relates to the following areas: Search-Based Software Testing (SBST), and LLM-based unit test generation. 

\subsection{Search Based Software Testing (SBST)}

SBST employs search algorithms to automatically generate test cases \cite{10440574, DBLP:journals/corr/abs-2111-05003, 10485640, 9685297}, which can greatly reduce the time developers spend on test case creation \cite{9407455}, while also generating boundary cases and exceptional inputs that are often challenging to identify manually. Various SBST-based tools and algorithms have been developed to generate test cases for programming languages such as Java, Python, and JavaScript, some of which support multiple testing objectives, including statement and branch coverage \cite{10.1145/3526072.3527528, 10.1145/3573074.3573102, 9476240, 10.1145/3526072.3527534}.

EvoSuite \cite{FSE21_ObjectSeed, GECCO21_Fitness, SBST21_competition} is a well-known tool based on SBST, which automatically generates JUnit test suites that maximize code coverage.
Randoop \cite{10.1145/1297846.1297902} is a feedback-directed random test generation tool that generates test cases by randomly combining previously executed statements that did not result in failures.
Algorithm MOSA (Multi-Objective Simulated Annealing) \cite{ulungu1999mosa}, and algorithms related to it, such as DynaMOSA \cite{panichella2017automated} and MIO (Many Independent Objective) \cite{arcuri2017many, arcuri2018test}, are used for test generation, and particularly effective at handling multiple objectives, containing line coverage, branch coverage, and multiple mutants in mutation testing.
Sapienz \cite{mao2016sapienz, arcuschin2023empirical, moreno2020algorithm} is an approach to Android testing that uses multi-objective SBST to optimize test sequences for brevity and effectiveness in revealing faults. Sapienze leverages a combination of random fuzzing, systematic exploration, and SBST. Pynguin is an extendable test-generation framework for Python, which is a dynamic type programming language \cite{lukasczyk2022pynguin, lukasczyk2023empirical, guerino2023experimental, salari2023automating}.

Despite the development of many SBST-based test case generation tools, they have notable limitations. These tools typically produce only boundary condition test cases, with assertions limited to simple equality checks. Additionally, when software updates occur, generating new test cases can be time-consuming, even for minimal changes. Furthermore, inaccurate type inference restricts their effectiveness, particularly in dynamically-typed languages like Python, which employs duck typing \cite{DBLP:journals/corr/abs-2111-05003}. In contrast, \toolname generates test cases providing more accurate type inference, enhancing overall test effectiveness.

\subsection{LLM-Based Unit Test Generation}




LLM-based unit test generation \cite{10329992, yang2024empirical} is the technique that leverages large language models trained to automatically generate unit tests for software code. There has been a large number of works or tools generating test cases with Large Language Models (LLMs) \cite{yang2024enhancing, xu2024mr, yang2024evaluation}, demonstrating impressive results. MuTAP \cite{dakhel2024effective} is a prompt-based learning technique to generate effective test cases with LLMs, which improves the effectiveness of test cases generated by LLMs in terms of revealing bugs by leveraging mutation testing. TestPilot \cite{Empirical_Evaluation} is a tool for automatically generating unit tests for npm packages written in JavaScript/TypeScript using LLM, which provides LLM with the signature and implementation of the function under test, along with usage extracted from the documentation. 
ChatUniTest \cite{chen2024chatunitest, xie2023chatunitest} utilizes an LLM-based approach encompassing valuable context in prompts and rectifying errors in generated unit tests.
ChatTester \cite{yuan2023no} is a Maven plugin similar to ChatUnitTest above, which leverages ChatGPT to improve the quality of its generated tests. LLM4Fin \cite{10.1145/3650212.3680388} is designed for testing real-world stock-trading software, which generates comprehensive testable scenarios for extracted business rules. 
CodaMosa \cite{Codamosa} developed by Microsoft combines SBST and LLMs, using LLM to help SBST's exploration. LLM will provide examples for SBST when its coverage improvements stall to help SBST search more useful areas.
Recent work \cite{zhang2025exploring} explores automated assertion generation with LLMs, conducting a large-scale evaluation of different models and demonstrating their effectiveness in improving assertion quality and detecting real-world bugs.
Reed et al. \cite{reed2025practical} investigates state-based testing techniques for object-oriented software, leveraging class-based unit tests and state-driven methodologies to improve testing effectiveness in large-scale systems.

While existing tools rely on prompts to guide LLMs in generating test cases, they fail to incorporate type information. For instance, TestPilot does not adapt its prompts based on type information, nor does it refine them when type information does not improve. 
Similarly, CodaMosa only prompts the LLM when necessary to support SBST, positioning the LLM as a supplementary component rather than a primary driver of test case generation. 
Type inference has achieved considerable success~\cite{yang2025runtyper}, yet its potential has not been effectively exploited in LLM-based test generation.
Pytlm~\cite{yang2025llm} assists SBST through inferred type information, yet does not leverage user-defined type information to improve test generation.
In contrast, our tool, \toolname, leverages type information to guide the LLM in generating test cases that are more likely to improve coverage. thereby enriching the LLM with additional, type-aware information.


\section{Conclusion}

We propose a type-aware approach to automated regression test generation, designed to enhance the validity and correctness of test cases for Python programs by inferring and incorporating precise type information during test construction.
\toolname leverages call graph-guided analysis to more effectively capture the semantic context of function parameters. 
Furthermore, by employing behavior-guided parameter type inference, \toolname improves the accuracy of parameter type prediction and strengthens the type soundness of generated test inputs.
To ensure robustness, \toolname integrates an automated fault repair mechanism, enabling the production of more comprehensive and reliable test suites. 
We conducted an extensive empirical evaluation across 183 real-world Python modules. 
The results show that \toolname consistently outperforms state-of-the-art baselines in both the effectiveness and stability of the generated test cases.
Future work will focus on extending the applicability of \toolname to other dynamically typed languages and further improving its efficiency.

\bibliographystyle{ACM-Reference-Format}
\bibliography{reference}


\begin{thebibliography}{62}


\ifx \showCODEN    \undefined \def \showCODEN     #1{\unskip}     \fi
\ifx \showISBNx    \undefined \def \showISBNx     #1{\unskip}     \fi
\ifx \showISBNxiii \undefined \def \showISBNxiii  #1{\unskip}     \fi
\ifx \showISSN     \undefined \def \showISSN      #1{\unskip}     \fi
\ifx \showLCCN     \undefined \def \showLCCN      #1{\unskip}     \fi
\ifx \shownote     \undefined \def \shownote      #1{#1}          \fi
\ifx \showarticletitle \undefined \def \showarticletitle #1{#1}   \fi
\ifx \showURL      \undefined \def \showURL       {\relax}        \fi
\providecommand\bibfield[2]{#2}
\providecommand\bibinfo[2]{#2}
\providecommand\natexlab[1]{#1}
\providecommand\showeprint[2][]{arXiv:#2}

\bibitem[Andrews et~al\mbox{.}(2011)]%
        {andrews2011genetic}
\bibfield{author}{\bibinfo{person}{J.~H. Andrews}, \bibinfo{person}{T. Menzies}, {and} \bibinfo{person}{F.~C.~H. Li}.} \bibinfo{year}{2011}\natexlab{}.
\newblock \showarticletitle{Genetic algorithms for randomized unit testing}.
\newblock \bibinfo{journal}{\emph{IEEE Transactions on Software Engineering}} \bibinfo{volume}{37}, \bibinfo{number}{1} (\bibinfo{year}{2011}), \bibinfo{pages}{80--94}.
\newblock


\bibitem[Arcuri(2017)]%
        {arcuri2017many}
\bibfield{author}{\bibinfo{person}{Andrea Arcuri}.} \bibinfo{year}{2017}\natexlab{}.
\newblock \showarticletitle{Many independent objective (MIO) algorithm for test suite generation}. In \bibinfo{booktitle}{\emph{Search Based Software Engineering: 9th International Symposium, SSBSE 2017, Paderborn, Germany, September 9-11, 2017, Proceedings 9}}. Springer, \bibinfo{pages}{3--17}.
\newblock


\bibitem[Arcuri(2018)]%
        {arcuri2018test}
\bibfield{author}{\bibinfo{person}{Andrea Arcuri}.} \bibinfo{year}{2018}\natexlab{}.
\newblock \showarticletitle{Test suite generation with the Many Independent Objective (MIO) algorithm}.
\newblock \bibinfo{journal}{\emph{Information and Software Technology}}  \bibinfo{volume}{104} (\bibinfo{year}{2018}), \bibinfo{pages}{195--206}.
\newblock


\bibitem[Arcuschin et~al\mbox{.}(2023)]%
        {arcuschin2023empirical}
\bibfield{author}{\bibinfo{person}{Iv{\'a}n Arcuschin}, \bibinfo{person}{Juan~Pablo Galeotti}, {and} \bibinfo{person}{Diego Garbervetsky}.} \bibinfo{year}{2023}\natexlab{}.
\newblock \showarticletitle{An Empirical Study on How Sapienz Achieves Coverage and Crash Detection}.
\newblock \bibinfo{journal}{\emph{Journal of Software: Evolution and Process}} \bibinfo{volume}{35}, \bibinfo{number}{4} (\bibinfo{year}{2023}), \bibinfo{pages}{e2411}.
\newblock


\bibitem[Chen et~al\mbox{.}(2021)]%
        {chen2021evaluating}
\bibfield{author}{\bibinfo{person}{Mark Chen}, \bibinfo{person}{Jerry Tworek}, \bibinfo{person}{Heewoo Jun}, \bibinfo{person}{Qiming Yuan}, \bibinfo{person}{Henrique Ponde De~Oliveira Pinto}, \bibinfo{person}{Jared Kaplan}, \bibinfo{person}{Harri Edwards}, \bibinfo{person}{Yuri Burda}, \bibinfo{person}{Nicholas Joseph}, \bibinfo{person}{Greg Brockman}, {et~al\mbox{.}}} \bibinfo{year}{2021}\natexlab{}.
\newblock \showarticletitle{Evaluating large language models trained on code}.
\newblock \bibinfo{journal}{\emph{arXiv preprint arXiv:2107.03374}} (\bibinfo{year}{2021}).
\newblock


\bibitem[Chen et~al\mbox{.}(2024)]%
        {chen2024chatunitest}
\bibfield{author}{\bibinfo{person}{Yinghao Chen}, \bibinfo{person}{Zehao Hu}, \bibinfo{person}{Chen Zhi}, \bibinfo{person}{Junxiao Han}, \bibinfo{person}{Shuiguang Deng}, {and} \bibinfo{person}{Jianwei Yin}.} \bibinfo{year}{2024}\natexlab{}.
\newblock \showarticletitle{Chatunitest: A framework for llm-based test generation}. In \bibinfo{booktitle}{\emph{Companion Proceedings of the 32nd ACM International Conference on the Foundations of Software Engineering}}. \bibinfo{pages}{572--576}.
\newblock


\bibitem[Dakhel et~al\mbox{.}(2024)]%
        {dakhel2024effective}
\bibfield{author}{\bibinfo{person}{Arghavan~Moradi Dakhel}, \bibinfo{person}{Amin Nikanjam}, \bibinfo{person}{Vahid Majdinasab}, \bibinfo{person}{Foutse Khomh}, {and} \bibinfo{person}{Michel~C Desmarais}.} \bibinfo{year}{2024}\natexlab{}.
\newblock \showarticletitle{Effective test generation using pre-trained large language models and mutation testing}.
\newblock \bibinfo{journal}{\emph{Information and Software Technology}}  \bibinfo{volume}{171} (\bibinfo{year}{2024}), \bibinfo{pages}{107468}.
\newblock


\bibitem[Davis et~al\mbox{.}(2023)]%
        {davis2023nanofuzz}
\bibfield{author}{\bibinfo{person}{Matthew~C Davis}, \bibinfo{person}{Sangheon Choi}, \bibinfo{person}{Sam Estep}, \bibinfo{person}{Brad~A Myers}, {and} \bibinfo{person}{Joshua Sunshine}.} \bibinfo{year}{2023}\natexlab{}.
\newblock \showarticletitle{NaNofuzz: A Usable Tool for Automatic Test Generation}. In \bibinfo{booktitle}{\emph{Proceedings of the 31st ACM Joint European Software Engineering Conference and Symposium on the Foundations of Software Engineering}}. \bibinfo{pages}{1114--1126}.
\newblock


\bibitem[Derakhshanfar and Devroey(2023)]%
        {10.1145/3526072.3527528}
\bibfield{author}{\bibinfo{person}{Pouria Derakhshanfar} {and} \bibinfo{person}{Xavier Devroey}.} \bibinfo{year}{2023}\natexlab{}.
\newblock \showarticletitle{Basic block coverage for unit test generation at the SBST 2022 tool competition}. In \bibinfo{booktitle}{\emph{Proceedings of the 15th Workshop on Search-Based Software Testing}} (Pittsburgh, Pennsylvania) \emph{(\bibinfo{series}{SBST '22})}. \bibinfo{publisher}{Association for Computing Machinery}, \bibinfo{address}{New York, NY, USA}, \bibinfo{pages}{37–38}.
\newblock
\showISBNx{9781450393188}
\href{https://doi.org/10.1145/3526072.3527528}{doi:\nolinkurl{10.1145/3526072.3527528}}


\bibitem[Eibich et~al\mbox{.}(2024)]%
        {eibich2024aragog}
\bibfield{author}{\bibinfo{person}{Matou{\v{s}} Eibich}, \bibinfo{person}{Shivay Nagpal}, {and} \bibinfo{person}{Alexander Fred-Ojala}.} \bibinfo{year}{2024}\natexlab{}.
\newblock \showarticletitle{ARAGOG: Advanced RAG output grading}.
\newblock \bibinfo{journal}{\emph{arXiv preprint arXiv:2404.01037}} (\bibinfo{year}{2024}).
\newblock


\bibitem[Ferdous et~al\mbox{.}(2023)]%
        {10.1145/3526072.3527534}
\bibfield{author}{\bibinfo{person}{Raihana Ferdous}, \bibinfo{person}{Chia-kang Hung}, \bibinfo{person}{Fitsum Kifetew}, \bibinfo{person}{Davide Prandi}, {and} \bibinfo{person}{Angelo Susi}.} \bibinfo{year}{2023}\natexlab{}.
\newblock \showarticletitle{EvoMBT at the SBST 2022 tool competition}. In \bibinfo{booktitle}{\emph{Proceedings of the 15th Workshop on Search-Based Software Testing}} (Pittsburgh, Pennsylvania) \emph{(\bibinfo{series}{SBST '22})}. \bibinfo{publisher}{Association for Computing Machinery}, \bibinfo{address}{New York, NY, USA}, \bibinfo{pages}{51–52}.
\newblock
\showISBNx{9781450393188}
\href{https://doi.org/10.1145/3526072.3527534}{doi:\nolinkurl{10.1145/3526072.3527534}}


\bibitem[Gao(2023)]%
        {gao2023prompt}
\bibfield{author}{\bibinfo{person}{Andrew Gao}.} \bibinfo{year}{2023}\natexlab{}.
\newblock \showarticletitle{Prompt engineering for large language models}.
\newblock \bibinfo{journal}{\emph{Available at SSRN 4504303}} (\bibinfo{year}{2023}).
\newblock


\bibitem[Gargari and Keyvanpour(2021)]%
        {9685297}
\bibfield{author}{\bibinfo{person}{Sepideh~Kashefi Gargari} {and} \bibinfo{person}{Mohammd~Reza Keyvanpour}.} \bibinfo{year}{2021}\natexlab{}.
\newblock \showarticletitle{SBST challenges from the perspective of the test techniques}. In \bibinfo{booktitle}{\emph{2021 12th International Conference on Information and Knowledge Technology (IKT)}}. \bibinfo{pages}{119--123}.
\newblock
\href{https://doi.org/10.1109/IKT54664.2021.9685297}{doi:\nolinkurl{10.1109/IKT54664.2021.9685297}}


\bibitem[Godefroid et~al\mbox{.}(2005)]%
        {godefroid2005dart}
\bibfield{author}{\bibinfo{person}{Patrice Godefroid}, \bibinfo{person}{Nils Klarlund}, {and} \bibinfo{person}{Koushik Sen}.} \bibinfo{year}{2005}\natexlab{}.
\newblock \showarticletitle{{DART: Directed automated random testing}}. In \bibinfo{booktitle}{\emph{Proceedings of the 2005 ACM SIGPLAN conference on Programming language design and implementation}}. ACM, \bibinfo{pages}{213--223}.
\newblock


\bibitem[Guerino and Vincenzi(2023)]%
        {guerino2023experimental}
\bibfield{author}{\bibinfo{person}{Lucca Guerino} {and} \bibinfo{person}{Auri Vincenzi}.} \bibinfo{year}{2023}\natexlab{}.
\newblock \showarticletitle{An Experimental Study Evaluating Cost, Adequacy, and Effectiveness of Pynguin's Test Sets}. In \bibinfo{booktitle}{\emph{Proceedings of the 8th Brazilian Symposium on Systematic and Automated Software Testing}}. \bibinfo{pages}{5--14}.
\newblock


\bibitem[Guizzo and Panichella(2023)]%
        {10.1145/3573074.3573102}
\bibfield{author}{\bibinfo{person}{Giovani Guizzo} {and} \bibinfo{person}{Sebastiano Panichella}.} \bibinfo{year}{2023}\natexlab{}.
\newblock \showarticletitle{Fuzzing vs SBST: Intersections \& Differences}.
\newblock \bibinfo{journal}{\emph{SIGSOFT Softw. Eng. Notes}} \bibinfo{volume}{48}, \bibinfo{number}{1} (\bibinfo{date}{Jan.} \bibinfo{year}{2023}), \bibinfo{pages}{105–107}.
\newblock
\showISSN{0163-5948}
\href{https://doi.org/10.1145/3573074.3573102}{doi:\nolinkurl{10.1145/3573074.3573102}}


\bibitem[Hwang et~al\mbox{.}(2021)]%
        {hwang2021justgen}
\bibfield{author}{\bibinfo{person}{Sungjae Hwang}, \bibinfo{person}{Sungho Lee}, \bibinfo{person}{Jihoon Kim}, {and} \bibinfo{person}{Sukyoung Ryu}.} \bibinfo{year}{2021}\natexlab{}.
\newblock \showarticletitle{Justgen: Effective test generation for unspecified JNI behaviors on jvms}. In \bibinfo{booktitle}{\emph{2021 IEEE/ACM 43rd International Conference on Software Engineering (ICSE)}}. IEEE, \bibinfo{pages}{1708--1718}.
\newblock


\bibitem[Juan Altmayer~Pizzorno(2024)]%
        {CoverUp}
\bibfield{author}{\bibinfo{person}{Emery D.~Berger Juan Altmayer~Pizzorno}.} \bibinfo{year}{2024}\natexlab{}.
\newblock \showarticletitle{CoverUp: Coverage-Guided LLM-Based Test GenerationCoverUp: Coverage-Guided LLM-Based Test Generation}.
\newblock \bibinfo{journal}{\emph{arXiv preprint arXiv:2403.16218, 2024 - arxiv.org}} (\bibinfo{year}{2024}).
\newblock


\bibitem[Lemieux et~al\mbox{.}(2023)]%
        {Codamosa}
\bibfield{author}{\bibinfo{person}{Caroline Lemieux}, \bibinfo{person}{Jeevana~Priya Inala}, \bibinfo{person}{Shuvendu~K Lahiri}, {and} \bibinfo{person}{Siddhartha Sen}.} \bibinfo{year}{2023}\natexlab{}.
\newblock \showarticletitle{Codamosa: Escaping coverage plateaus in test generation with pre-trained large language models}. In \bibinfo{booktitle}{\emph{2023 IEEE/ACM 45th International Conference on Software Engineering (ICSE)}}. IEEE, \bibinfo{pages}{919--931}.
\newblock


\bibitem[Lin et~al\mbox{.}(2021)]%
        {FSE21_ObjectSeed}
\bibfield{author}{\bibinfo{person}{Yun Lin}, \bibinfo{person}{You~Sheng Ong}, \bibinfo{person}{Jun Sun}, \bibinfo{person}{Gordon Fraser}, {and} \bibinfo{person}{Jin~Song Dong}.} \bibinfo{year}{2021}\natexlab{}.
\newblock \showarticletitle{Graph-Based Seed Object Synthesis for Search-Based Unit Testing}. In \bibinfo{booktitle}{\emph{Proceedings of the 29th ACM Joint Meeting on European Software Engineering Conference and Symposium on the Foundations of Software Engineering}} (Athens, Greece) \emph{(\bibinfo{series}{ESEC/FSE 2021})}. \bibinfo{publisher}{Association for Computing Machinery}, \bibinfo{address}{New York, NY, USA}, \bibinfo{pages}{1068–1080}.
\newblock
\showISBNx{9781450385626}
\href{https://doi.org/10.1145/3468264.3468619}{doi:\nolinkurl{10.1145/3468264.3468619}}


\bibitem[Lukasczyk and Fraser(2022)]%
        {lukasczyk2022pynguin}
\bibfield{author}{\bibinfo{person}{Stephan Lukasczyk} {and} \bibinfo{person}{Gordon Fraser}.} \bibinfo{year}{2022}\natexlab{}.
\newblock \showarticletitle{Pynguin: Automated unit test generation for python}. In \bibinfo{booktitle}{\emph{Proceedings of the ACM/IEEE 44th International Conference on Software Engineering: Companion Proceedings}}. \bibinfo{pages}{168--172}.
\newblock


\bibitem[Lukasczyk et~al\mbox{.}(2021)]%
        {DBLP:journals/corr/abs-2111-05003}
\bibfield{author}{\bibinfo{person}{Stephan Lukasczyk}, \bibinfo{person}{Florian Kroi{\ss}}, {and} \bibinfo{person}{Gordon Fraser}.} \bibinfo{year}{2021}\natexlab{}.
\newblock \showarticletitle{An Empirical Study of Automated Unit Test Generation for Python}.
\newblock \bibinfo{journal}{\emph{CoRR}}  \bibinfo{volume}{abs/2111.05003} (\bibinfo{year}{2021}).
\newblock
\showeprint[arXiv]{2111.05003}
\urldef\tempurl%
\url{https://arxiv.org/abs/2111.05003}
\showURL{%
\tempurl}


\bibitem[Lukasczyk et~al\mbox{.}(2023)]%
        {lukasczyk2023empirical}
\bibfield{author}{\bibinfo{person}{Stephan Lukasczyk}, \bibinfo{person}{Florian Kroi{\ss}}, {and} \bibinfo{person}{Gordon Fraser}.} \bibinfo{year}{2023}\natexlab{}.
\newblock \showarticletitle{An empirical study of automated unit test generation for Python}.
\newblock \bibinfo{journal}{\emph{Empirical Software Engineering}} \bibinfo{volume}{28}, \bibinfo{number}{2} (\bibinfo{year}{2023}), \bibinfo{pages}{36}.
\newblock


\bibitem[Ma et~al\mbox{.}(2015)]%
        {ma2015grt}
\bibfield{author}{\bibinfo{person}{Lin Ma}, \bibinfo{person}{Cyril Artho}, \bibinfo{person}{Chao Zhang}, {et~al\mbox{.}}} \bibinfo{year}{2015}\natexlab{}.
\newblock \showarticletitle{{Grt: Program-analysis-guided random testing}}. In \bibinfo{booktitle}{\emph{2015 30th IEEE/ACM International Conference on Automated Software Engineering (ASE)}}. IEEE, \bibinfo{pages}{212--223}.
\newblock


\bibitem[Mao et~al\mbox{.}(2016)]%
        {mao2016sapienz}
\bibfield{author}{\bibinfo{person}{Ke Mao}, \bibinfo{person}{Mark Harman}, {and} \bibinfo{person}{Yue Jia}.} \bibinfo{year}{2016}\natexlab{}.
\newblock \showarticletitle{Sapienz: Multi-objective automated testing for android applications}. In \bibinfo{booktitle}{\emph{Proceedings of the 25th international symposium on software testing and analysis}}. \bibinfo{pages}{94--105}.
\newblock


\bibitem[Mao et~al\mbox{.}(2020)]%
        {mao2020multi}
\bibfield{author}{\bibinfo{person}{Yuning Mao}, \bibinfo{person}{Yanru Qu}, \bibinfo{person}{Yiqing Xie}, \bibinfo{person}{Xiang Ren}, {and} \bibinfo{person}{Jiawei Han}.} \bibinfo{year}{2020}\natexlab{}.
\newblock \showarticletitle{Multi-document summarization with maximal marginal relevance-guided reinforcement learning}.
\newblock \bibinfo{journal}{\emph{arXiv preprint arXiv:2010.00117}} (\bibinfo{year}{2020}).
\newblock


\bibitem[Mir et~al\mbox{.}(2021)]%
        {mir2021manytypes4py}
\bibfield{author}{\bibinfo{person}{Amir~M Mir}, \bibinfo{person}{Evaldas Lato{\v{s}}kinas}, {and} \bibinfo{person}{Georgios Gousios}.} \bibinfo{year}{2021}\natexlab{}.
\newblock \showarticletitle{Manytypes4py: A benchmark python dataset for machine learning-based type inference}. In \bibinfo{booktitle}{\emph{2021 IEEE/ACM 18th International Conference on Mining Software Repositories (MSR)}}. IEEE, \bibinfo{pages}{585--589}.
\newblock


\bibitem[Mir et~al\mbox{.}(2022)]%
        {mir2022type4py}
\bibfield{author}{\bibinfo{person}{Amir~M Mir}, \bibinfo{person}{Evaldas Lato{\v{s}}kinas}, \bibinfo{person}{Sebastian Proksch}, {and} \bibinfo{person}{Georgios Gousios}.} \bibinfo{year}{2022}\natexlab{}.
\newblock \showarticletitle{Type4py: Practical deep similarity learning-based type inference for python}. In \bibinfo{booktitle}{\emph{Proceedings of the 44th International Conference on Software Engineering}}. \bibinfo{pages}{2241--2252}.
\newblock


\bibitem[Moghadam et~al\mbox{.}(2021)]%
        {9476240}
\bibfield{author}{\bibinfo{person}{Mahshid~Helali Moghadam}, \bibinfo{person}{Markus Borg}, {and} \bibinfo{person}{Seyed~Jalaleddin Mousavirad}.} \bibinfo{year}{2021}\natexlab{}.
\newblock \showarticletitle{Deeper at the SBST 2021 Tool Competition: ADAS Testing Using Multi-Objective Search}. In \bibinfo{booktitle}{\emph{2021 IEEE/ACM 14th International Workshop on Search-Based Software Testing (SBST)}}. \bibinfo{pages}{40--41}.
\newblock
\href{https://doi.org/10.1109/SBST52555.2021.00018}{doi:\nolinkurl{10.1109/SBST52555.2021.00018}}


\bibitem[Moreno et~al\mbox{.}(2020)]%
        {moreno2020algorithm}
\bibfield{author}{\bibinfo{person}{Iv{\'a}n~Arcuschin Moreno}, \bibinfo{person}{Juan~Pablo Galeotti}, {and} \bibinfo{person}{Diego Garbervetsky}.} \bibinfo{year}{2020}\natexlab{}.
\newblock \showarticletitle{Algorithm or Representation? An empirical study on how SAPIENZ achieves coverage}. In \bibinfo{booktitle}{\emph{Proceedings of the IEEE/ACM 1st International Conference on Automation of Software Test}}. \bibinfo{pages}{61--70}.
\newblock


\bibitem[Oh and Oh(2022)]%
        {oh2022pyter}
\bibfield{author}{\bibinfo{person}{Wonseok Oh} {and} \bibinfo{person}{Hakjoo Oh}.} \bibinfo{year}{2022}\natexlab{}.
\newblock \showarticletitle{PyTER: effective program repair for Python type errors}. In \bibinfo{booktitle}{\emph{Proceedings of the 30th ACM Joint European Software Engineering Conference and Symposium on the Foundations of Software Engineering}}. \bibinfo{pages}{922--934}.
\newblock


\bibitem[Pacheco and Ernst(2007)]%
        {10.1145/1297846.1297902}
\bibfield{author}{\bibinfo{person}{Carlos Pacheco} {and} \bibinfo{person}{Michael~D. Ernst}.} \bibinfo{year}{2007}\natexlab{}.
\newblock \showarticletitle{Randoop: feedback-directed random testing for Java}. In \bibinfo{booktitle}{\emph{Companion to the 22nd ACM SIGPLAN Conference on Object-Oriented Programming Systems and Applications Companion}} (Montreal, Quebec, Canada) \emph{(\bibinfo{series}{OOPSLA '07})}. \bibinfo{publisher}{Association for Computing Machinery}, \bibinfo{address}{New York, NY, USA}, \bibinfo{pages}{815–816}.
\newblock
\showISBNx{9781595938657}
\href{https://doi.org/10.1145/1297846.1297902}{doi:\nolinkurl{10.1145/1297846.1297902}}


\bibitem[Panichella et~al\mbox{.}(2017)]%
        {panichella2017automated}
\bibfield{author}{\bibinfo{person}{Annibale Panichella}, \bibinfo{person}{Fitsum~Meshesha Kifetew}, {and} \bibinfo{person}{Paolo Tonella}.} \bibinfo{year}{2017}\natexlab{}.
\newblock \showarticletitle{Automated test case generation as a many-objective optimisation problem with dynamic selection of the targets}.
\newblock \bibinfo{journal}{\emph{IEEE Transactions on Software Engineering}} \bibinfo{volume}{44}, \bibinfo{number}{2} (\bibinfo{year}{2017}), \bibinfo{pages}{122--158}.
\newblock


\bibitem[Peng et~al\mbox{.}(2022)]%
        {peng2022static}
\bibfield{author}{\bibinfo{person}{Yun Peng}, \bibinfo{person}{Cuiyun Gao}, \bibinfo{person}{Zongjie Li}, \bibinfo{person}{Bowei Gao}, \bibinfo{person}{David Lo}, \bibinfo{person}{Qirun Zhang}, {and} \bibinfo{person}{Michael Lyu}.} \bibinfo{year}{2022}\natexlab{}.
\newblock \showarticletitle{Static inference meets deep learning: a hybrid type inference approach for python}. In \bibinfo{booktitle}{\emph{Proceedings of the 44th International Conference on Software Engineering}}. \bibinfo{pages}{2019--2030}.
\newblock


\bibitem[Peng et~al\mbox{.}(2023)]%
        {peng2023generative}
\bibfield{author}{\bibinfo{person}{Yun Peng}, \bibinfo{person}{Chaozheng Wang}, \bibinfo{person}{Wenxuan Wang}, \bibinfo{person}{Cuiyun Gao}, {and} \bibinfo{person}{Michael~R Lyu}.} \bibinfo{year}{2023}\natexlab{}.
\newblock \showarticletitle{Generative type inference for python}. In \bibinfo{booktitle}{\emph{2023 38th IEEE/ACM International Conference on Automated Software Engineering (ASE)}}. IEEE, \bibinfo{pages}{988--999}.
\newblock


\bibitem[Reed et~al\mbox{.}(2025)]%
        {reed2025practical}
\bibfield{author}{\bibinfo{person}{HG Reed}, \bibinfo{person}{CD Turner}, \bibinfo{person}{JB Aibel}, {and} \bibinfo{person}{JT Dalton}.} \bibinfo{year}{2025}\natexlab{}.
\newblock \showarticletitle{Practical object-oriented state-based unit testing}.
\newblock \bibinfo{journal}{\emph{WIT Transactions on Information and Communication Technologies}}  \bibinfo{volume}{9} (\bibinfo{year}{2025}).
\newblock


\bibitem[Salari et~al\mbox{.}(2023)]%
        {salari2023automating}
\bibfield{author}{\bibinfo{person}{Mikael~Ebrahimi Salari}, \bibinfo{person}{Eduard~Paul Enoiu}, \bibinfo{person}{Cristina Seceleanu}, \bibinfo{person}{Wasif Afzal}, {and} \bibinfo{person}{Filip Sebek}.} \bibinfo{year}{2023}\natexlab{}.
\newblock \showarticletitle{Automating test generation of industrial control software through a plc-to-python translation framework and pynguin}. In \bibinfo{booktitle}{\emph{2023 30th Asia-Pacific Software Engineering Conference (APSEC)}}. IEEE, \bibinfo{pages}{431--440}.
\newblock


\bibitem[Salis et~al\mbox{.}(2021)]%
        {salis2021pycg}
\bibfield{author}{\bibinfo{person}{Vitalis Salis}, \bibinfo{person}{Thodoris Sotiropoulos}, \bibinfo{person}{Panos Louridas}, \bibinfo{person}{Diomidis Spinellis}, {and} \bibinfo{person}{Dimitris Mitropoulos}.} \bibinfo{year}{2021}\natexlab{}.
\newblock \showarticletitle{Pycg: Practical call graph generation in python}. In \bibinfo{booktitle}{\emph{2021 IEEE/ACM 43rd International Conference on Software Engineering (ICSE)}}. IEEE, \bibinfo{pages}{1646--1657}.
\newblock


\bibitem[Sch{\"a}fer et~al\mbox{.}(2023)]%
        {Empirical_Evaluation}
\bibfield{author}{\bibinfo{person}{Max Sch{\"a}fer}, \bibinfo{person}{Sarah Nadi}, \bibinfo{person}{Aryaz Eghbali}, {and} \bibinfo{person}{Frank Tip}.} \bibinfo{year}{2023}\natexlab{}.
\newblock \showarticletitle{An empirical evaluation of using large language models for automated unit test generation}.
\newblock \bibinfo{journal}{\emph{IEEE Transactions on Software Engineering}} (\bibinfo{year}{2023}).
\newblock


\bibitem[Schäfer et~al\mbox{.}(2024)]%
        {10329992}
\bibfield{author}{\bibinfo{person}{Max Schäfer}, \bibinfo{person}{Sarah Nadi}, \bibinfo{person}{Aryaz Eghbali}, {and} \bibinfo{person}{Frank Tip}.} \bibinfo{year}{2024}\natexlab{}.
\newblock \showarticletitle{An Empirical Evaluation of Using Large Language Models for Automated Unit Test Generation}.
\newblock \bibinfo{journal}{\emph{IEEE Transactions on Software Engineering}} \bibinfo{volume}{50}, \bibinfo{number}{1} (\bibinfo{year}{2024}), \bibinfo{pages}{85--105}.
\newblock
\href{https://doi.org/10.1109/TSE.2023.3334955}{doi:\nolinkurl{10.1109/TSE.2023.3334955}}


\bibitem[Tang et~al\mbox{.}(2024)]%
        {10485640}
\bibfield{author}{\bibinfo{person}{Yutian Tang}, \bibinfo{person}{Zhijie Liu}, \bibinfo{person}{Zhichao Zhou}, {and} \bibinfo{person}{Xiapu Luo}.} \bibinfo{year}{2024}\natexlab{}.
\newblock \showarticletitle{ChatGPT vs SBST: A Comparative Assessment of Unit Test Suite Generation}.
\newblock \bibinfo{journal}{\emph{IEEE Transactions on Software Engineering}} \bibinfo{volume}{50}, \bibinfo{number}{6} (\bibinfo{year}{2024}), \bibinfo{pages}{1340--1359}.
\newblock
\href{https://doi.org/10.1109/TSE.2024.3382365}{doi:\nolinkurl{10.1109/TSE.2024.3382365}}


\bibitem[Ulungu et~al\mbox{.}(1999)]%
        {ulungu1999mosa}
\bibfield{author}{\bibinfo{person}{Ekunda~Lukata Ulungu}, \bibinfo{person}{JFPH Teghem}, \bibinfo{person}{PH Fortemps}, {and} \bibinfo{person}{Daniel Tuyttens}.} \bibinfo{year}{1999}\natexlab{}.
\newblock \showarticletitle{MOSA method: a tool for solving multiobjective combinatorial optimization problems}.
\newblock \bibinfo{journal}{\emph{Journal of multicriteria decision analysis}} \bibinfo{volume}{8}, \bibinfo{number}{4} (\bibinfo{year}{1999}), \bibinfo{pages}{221}.
\newblock


\bibitem[Vasudevan et~al\mbox{.}(2021)]%
        {9407455}
\bibfield{author}{\bibinfo{person}{M~S Vasudevan}, \bibinfo{person}{Santosh Biswas}, {and} \bibinfo{person}{Aryabartta Sahu}.} \bibinfo{year}{2021}\natexlab{}.
\newblock \showarticletitle{Automated Low-Cost SBST Optimization Techniques for Processor Testing}. In \bibinfo{booktitle}{\emph{2021 34th International Conference on VLSI Design and 2021 20th International Conference on Embedded Systems (VLSID)}}. \bibinfo{pages}{299--304}.
\newblock
\href{https://doi.org/10.1109/VLSID51830.2021.00056}{doi:\nolinkurl{10.1109/VLSID51830.2021.00056}}


\bibitem[Vogl et~al\mbox{.}(2021a)]%
        {GECCO21_Fitness}
\bibfield{author}{\bibinfo{person}{Sebastian Vogl}, \bibinfo{person}{Sebastian Schweikl}, {and} \bibinfo{person}{Gordon Fraser}.} \bibinfo{year}{2021}\natexlab{a}.
\newblock \showarticletitle{Encoding the certainty of boolean variables to improve the guidance for search-based test generation}. In \bibinfo{booktitle}{\emph{Proceedings of the Genetic and Evolutionary Computation Conference}}. \bibinfo{pages}{1088--1096}.
\newblock


\bibitem[Vogl et~al\mbox{.}(2021b)]%
        {SBST21_competition}
\bibfield{author}{\bibinfo{person}{Sebastian Vogl}, \bibinfo{person}{Sebastian Schweikl}, \bibinfo{person}{Gordon Fraser}, \bibinfo{person}{Andrea Arcuri}, \bibinfo{person}{Jose Campos}, {and} \bibinfo{person}{Annibale Panichella}.} \bibinfo{year}{2021}\natexlab{b}.
\newblock \showarticletitle{EVOSUITE at the SBST 2021 Tool Competition}. In \bibinfo{booktitle}{\emph{2021 IEEE/ACM 14th International Workshop on Search-Based Software Testing (SBST)}}. IEEE, \bibinfo{pages}{28--29}.
\newblock


\bibitem[Wang et~al\mbox{.}(2024)]%
        {10440574}
\bibfield{author}{\bibinfo{person}{Junjie Wang}, \bibinfo{person}{Yuchao Huang}, \bibinfo{person}{Chunyang Chen}, \bibinfo{person}{Zhe Liu}, \bibinfo{person}{Song Wang}, {and} \bibinfo{person}{Qing Wang}.} \bibinfo{year}{2024}\natexlab{}.
\newblock \showarticletitle{Software Testing With Large Language Models: Survey, Landscape, and Vision}.
\newblock \bibinfo{journal}{\emph{IEEE Transactions on Software Engineering}} \bibinfo{volume}{50}, \bibinfo{number}{4} (\bibinfo{year}{2024}), \bibinfo{pages}{911--936}.
\newblock
\href{https://doi.org/10.1109/TSE.2024.3368208}{doi:\nolinkurl{10.1109/TSE.2024.3368208}}


\bibitem[Wei et~al\mbox{.}(2022)]%
        {wei2022free}
\bibfield{author}{\bibinfo{person}{Anjiang Wei}, \bibinfo{person}{Yinlin Deng}, \bibinfo{person}{Chenyuan Yang}, {and} \bibinfo{person}{Lingming Zhang}.} \bibinfo{year}{2022}\natexlab{}.
\newblock \showarticletitle{Free lunch for testing: Fuzzing deep-learning libraries from open source}. In \bibinfo{booktitle}{\emph{Proceedings of the 44th International Conference on Software Engineering}}. \bibinfo{pages}{995--1007}.
\newblock


\bibitem[White et~al\mbox{.}(2023)]%
        {white2023prompt}
\bibfield{author}{\bibinfo{person}{Jules White}, \bibinfo{person}{Quchen Fu}, \bibinfo{person}{Sam Hays}, \bibinfo{person}{Michael Sandborn}, \bibinfo{person}{Carlos Olea}, \bibinfo{person}{Henry Gilbert}, \bibinfo{person}{Ashraf Elnashar}, \bibinfo{person}{Jesse Spencer-Smith}, {and} \bibinfo{person}{Douglas~C Schmidt}.} \bibinfo{year}{2023}\natexlab{}.
\newblock \showarticletitle{A prompt pattern catalog to enhance prompt engineering with chatgpt}.
\newblock \bibinfo{journal}{\emph{arXiv preprint arXiv:2302.11382}} (\bibinfo{year}{2023}).
\newblock


\bibitem[Widyasari et~al\mbox{.}(2020)]%
        {widyasari2020bugsinpy}
\bibfield{author}{\bibinfo{person}{Ratnadira Widyasari}, \bibinfo{person}{Sheng~Qin Sim}, \bibinfo{person}{Camellia Lok}, \bibinfo{person}{Haodi Qi}, \bibinfo{person}{Jack Phan}, \bibinfo{person}{Qijin Tay}, \bibinfo{person}{Constance Tan}, \bibinfo{person}{Fiona Wee}, \bibinfo{person}{Jodie~Ethelda Tan}, \bibinfo{person}{Yuheng Yieh}, {et~al\mbox{.}}} \bibinfo{year}{2020}\natexlab{}.
\newblock \showarticletitle{Bugsinpy: a database of existing bugs in python programs to enable controlled testing and debugging studies}. In \bibinfo{booktitle}{\emph{Proceedings of the 28th ACM joint meeting on european software engineering conference and symposium on the foundations of software engineering}}. \bibinfo{pages}{1556--1560}.
\newblock


\bibitem[Xie et~al\mbox{.}(2023)]%
        {xie2023chatunitest}
\bibfield{author}{\bibinfo{person}{Zhuokui Xie}, \bibinfo{person}{Yinghao Chen}, \bibinfo{person}{Chen Zhi}, \bibinfo{person}{Shuiguang Deng}, {and} \bibinfo{person}{Jianwei Yin}.} \bibinfo{year}{2023}\natexlab{}.
\newblock \showarticletitle{ChatUniTest: a ChatGPT-based automated unit test generation tool}.
\newblock \bibinfo{journal}{\emph{arXiv preprint arXiv:2305.04764}} (\bibinfo{year}{2023}).
\newblock


\bibitem[Xu et~al\mbox{.}(2024)]%
        {xu2024mr}
\bibfield{author}{\bibinfo{person}{Congying Xu}, \bibinfo{person}{Songqiang Chen}, \bibinfo{person}{Jiarong Wu}, \bibinfo{person}{Shing-Chi Cheung}, \bibinfo{person}{Valerio Terragni}, \bibinfo{person}{Hengcheng Zhu}, {and} \bibinfo{person}{Jialun Cao}.} \bibinfo{year}{2024}\natexlab{}.
\newblock \showarticletitle{MR-Adopt: Automatic Deduction of Input Transformation Function for Metamorphic Testing}.
\newblock \bibinfo{journal}{\emph{arXiv preprint arXiv:2408.15815}} (\bibinfo{year}{2024}).
\newblock


\bibitem[Xue et~al\mbox{.}(2024)]%
        {10.1145/3650212.3680388}
\bibfield{author}{\bibinfo{person}{Zhiyi Xue}, \bibinfo{person}{Liangguo Li}, \bibinfo{person}{Senyue Tian}, \bibinfo{person}{Xiaohong Chen}, \bibinfo{person}{Pingping Li}, \bibinfo{person}{Liangyu Chen}, \bibinfo{person}{Tingting Jiang}, {and} \bibinfo{person}{Min Zhang}.} \bibinfo{year}{2024}\natexlab{}.
\newblock \showarticletitle{LLM4Fin: Fully Automating LLM-Powered Test Case Generation for FinTech Software Acceptance Testing}. In \bibinfo{booktitle}{\emph{Proceedings of the 33rd ACM SIGSOFT International Symposium on Software Testing and Analysis}} (Vienna, Austria) \emph{(\bibinfo{series}{ISSTA 2024})}. \bibinfo{publisher}{Association for Computing Machinery}, \bibinfo{address}{New York, NY, USA}, \bibinfo{pages}{1643–1655}.
\newblock
\showISBNx{9798400706127}
\href{https://doi.org/10.1145/3650212.3680388}{doi:\nolinkurl{10.1145/3650212.3680388}}


\bibitem[Yang et~al\mbox{.}(2024a)]%
        {yang2024enhancing}
\bibfield{author}{\bibinfo{person}{Chen Yang}, \bibinfo{person}{Junjie Chen}, \bibinfo{person}{Bin Lin}, \bibinfo{person}{Jianyi Zhou}, {and} \bibinfo{person}{Ziqi Wang}.} \bibinfo{year}{2024}\natexlab{a}.
\newblock \showarticletitle{Enhancing LLM-based Test Generation for Hard-to-Cover Branches via Program Analysis}.
\newblock \bibinfo{journal}{\emph{arXiv preprint arXiv:2404.04966}} (\bibinfo{year}{2024}).
\newblock


\bibitem[Yang et~al\mbox{.}(2025a)]%
        {yang2025runtyper}
\bibfield{author}{\bibinfo{person}{Guowei Yang}, \bibinfo{person}{Shilin He}, \bibinfo{person}{Fu Song}, {and} \bibinfo{person}{Yuqi Chen}.} \bibinfo{year}{2025}\natexlab{a}.
\newblock \showarticletitle{RunTyper: Enhancing Deep Type Inference Using Dynamic Analysis for Python}.
\newblock \bibinfo{journal}{\emph{ACM Transactions on Software Engineering and Methodology}} (\bibinfo{year}{2025}).
\newblock


\bibitem[Yang et~al\mbox{.}(2024b)]%
        {yang2024empirical}
\bibfield{author}{\bibinfo{person}{Lin Yang}, \bibinfo{person}{Chen Yang}, \bibinfo{person}{Shutao Gao}, \bibinfo{person}{Weijing Wang}, \bibinfo{person}{Bo Wang}, \bibinfo{person}{Qihao Zhu}, \bibinfo{person}{Xiao Chu}, \bibinfo{person}{Jianyi Zhou}, \bibinfo{person}{Guangtai Liang}, \bibinfo{person}{Qianxiang Wang}, {et~al\mbox{.}}} \bibinfo{year}{2024}\natexlab{b}.
\newblock \showarticletitle{An empirical study of unit test generation with large language models}.
\newblock \bibinfo{journal}{\emph{arXiv preprint arXiv:2406.18181}} (\bibinfo{year}{2024}).
\newblock


\bibitem[Yang et~al\mbox{.}(2024c)]%
        {yang2024evaluation}
\bibfield{author}{\bibinfo{person}{Lin Yang}, \bibinfo{person}{Chen Yang}, \bibinfo{person}{Shutao Gao}, \bibinfo{person}{Weijing Wang}, \bibinfo{person}{Bo Wang}, \bibinfo{person}{Qihao Zhu}, \bibinfo{person}{Xiao Chu}, \bibinfo{person}{Jianyi Zhou}, \bibinfo{person}{Guangtai Liang}, \bibinfo{person}{Qianxiang Wang}, {et~al\mbox{.}}} \bibinfo{year}{2024}\natexlab{c}.
\newblock \showarticletitle{On the Evaluation of Large Language Models in Unit Test Generation}. In \bibinfo{booktitle}{\emph{Proceedings of the 39th IEEE/ACM International Conference on Automated Software Engineering}}. \bibinfo{pages}{1607--1619}.
\newblock


\bibitem[Yang et~al\mbox{.}(2006)]%
        {Survey_of_coverage}
\bibfield{author}{\bibinfo{person}{Qian Yang}, \bibinfo{person}{J~Jenny Li}, {and} \bibinfo{person}{David Weiss}.} \bibinfo{year}{2006}\natexlab{}.
\newblock \showarticletitle{A survey of coverage based testing tools}. In \bibinfo{booktitle}{\emph{Proceedings of the 2006 international workshop on Automation of software test}}. \bibinfo{pages}{99--103}.
\newblock


\bibitem[Yang et~al\mbox{.}(2025b)]%
        {yang2025llm}
\bibfield{author}{\bibinfo{person}{Ruofan Yang}, \bibinfo{person}{Xianghua Xu}, {and} \bibinfo{person}{Ran Wang}.} \bibinfo{year}{2025}\natexlab{b}.
\newblock \showarticletitle{LLM-enhanced evolutionary test generation for untyped languages}.
\newblock \bibinfo{journal}{\emph{Automated Software Engineering}} \bibinfo{volume}{32}, \bibinfo{number}{1} (\bibinfo{year}{2025}), \bibinfo{pages}{20}.
\newblock


\bibitem[Yuan et~al\mbox{.}(2023)]%
        {yuan2023no}
\bibfield{author}{\bibinfo{person}{Zhiqiang Yuan}, \bibinfo{person}{Yiling Lou}, \bibinfo{person}{Mingwei Liu}, \bibinfo{person}{Shiji Ding}, \bibinfo{person}{Kaixin Wang}, \bibinfo{person}{Yixuan Chen}, {and} \bibinfo{person}{Xin Peng}.} \bibinfo{year}{2023}\natexlab{}.
\newblock \showarticletitle{No more manual tests? evaluating and improving chatgpt for unit test generation}.
\newblock \bibinfo{journal}{\emph{arXiv preprint arXiv:2305.04207}} (\bibinfo{year}{2023}).
\newblock


\bibitem[Zhang et~al\mbox{.}(2025)]%
        {zhang2025exploring}
\bibfield{author}{\bibinfo{person}{Quanjun Zhang}, \bibinfo{person}{Weifeng Sun}, \bibinfo{person}{Chunrong Fang}, \bibinfo{person}{Bowen Yu}, \bibinfo{person}{Hongyan Li}, \bibinfo{person}{Meng Yan}, \bibinfo{person}{Jianyi Zhou}, {and} \bibinfo{person}{Zhenyu Chen}.} \bibinfo{year}{2025}\natexlab{}.
\newblock \showarticletitle{Exploring automated assertion generation via large language models}.
\newblock \bibinfo{journal}{\emph{ACM Transactions on Software Engineering and Methodology}} \bibinfo{volume}{34}, \bibinfo{number}{3} (\bibinfo{year}{2025}), \bibinfo{pages}{1--25}.
\newblock


\bibitem[Zhang and Mesbah(2015)]%
        {zhang2015assertions}
\bibfield{author}{\bibinfo{person}{Yucheng Zhang} {and} \bibinfo{person}{Ali Mesbah}.} \bibinfo{year}{2015}\natexlab{}.
\newblock \showarticletitle{Assertions are strongly correlated with test suite effectiveness}. In \bibinfo{booktitle}{\emph{Proceedings of the 2015 10th Joint Meeting on Foundations of Software Engineering}}. \bibinfo{pages}{214--224}.
\newblock


\bibitem[Zhou et~al\mbox{.}(2022)]%
        {zhou2022selectively}
\bibfield{author}{\bibinfo{person}{Zhichao Zhou}, \bibinfo{person}{Yuming Zhou}, \bibinfo{person}{Chunrong Fang}, \bibinfo{person}{Zhenyu Chen}, {and} \bibinfo{person}{Yutian Tang}.} \bibinfo{year}{2022}\natexlab{}.
\newblock \showarticletitle{Selectively combining multiple coverage goals in search-based unit test generation}. In \bibinfo{booktitle}{\emph{Proceedings of the 37th IEEE/ACM International Conference on Automated Software Engineering}}. \bibinfo{pages}{1--12}.
\newblock


\end{thebibliography}

\appendix

\end{document}